\documentclass[floatfix,reprint,amsmath,amssymb,prapplied,superscriptaddress,aps]{revtex4-2}

\usepackage[dvipsnames]{xcolor}
\usepackage{graphicx}
\usepackage{amsmath,amssymb}
\usepackage{dcolumn}
\usepackage[utf8]{inputenc}
\usepackage[T1]{fontenc}
\usepackage{bm}
\usepackage{fontawesome5}
\usepackage{siunitx}
\usepackage{hyperref}

\definecolor{cerclegreen}{HTML}{139F46}
\definecolor{darkorange}{rgb}{1.0, 0.55, 0.0}
\newcommand{\jonathan}[1]{{ \color{black} #1}}

\usepackage{soul}

\hypersetup{
colorlinks=true, breaklinks=true, 
urlcolor=RoyalBlue, linkcolor=RoyalBlue, citecolor=RoyalBlue, 
pdftitle={},
pdfauthor={\textcopyright}, 
pdfsubject={},
pdfkeywords={}, 
pdfcreator={pdfLaTeX},
pdfproducer={LaTeX}
}

\begin{document}


\title{Nanoscale mapping of internal magnetization dynamics reveals how disorder shapes heat generation in magnetic particle hyperthermia }

\author{Elizabeth M. Jefremovas}
\email{elizabeth.jefremovas@.uni.lu}
\affiliation{Department of Physics and Materials Science, University of Luxembourg, 162A Avenue de la Faiencerie, L-1511 Luxembourg, Grand Duchy of Luxembourg}
\affiliation{Institute for Advanced Studies, University of Luxembourg, Campus Belval, L-4365 Esch-sur-Alzette, Luxembourg}
\author{Pauline Rooms}
\affiliation{DyNaMat, Department of Solid State Sciences, Ghent University, 9000 Ghent, Belgium}
\author{\'Alvaro Gallo-C\'ordova}
\affiliation{Instituto de Ciencia de Materiales de Madrid (ICMM-CSIC), Madrid 28049, Spain}
\author{Mar\'ia P. Morales}
\affiliation{Instituto de Ciencia de Materiales de Madrid (ICMM-CSIC), Madrid 28049, Spain}
\author{Frank Wiekhorst}
\affiliation{Physikalisch-Technische Bundesanstalt, D-10587 Berlin, Germany}
\author{Andreas Michels}
\affiliation{Department of Physics and Materials Science, University of Luxembourg, 162A Avenue de la Faiencerie, L-1511 Luxembourg, Grand Duchy of Luxembourg}
\author{Jonathan Leliaert}
\affiliation{DyNaMat, Department of Solid State Sciences, Ghent University, 9000 Ghent, Belgium}

\date{\today}

\begin{abstract}
Magnetic particle hyperthermia relies on the efficient conversion of magnetic field energy into heat in biomedical applications, yet the microscopic mechanisms governing heat generation within individual particles remain poorly understood. In this study, AC magnetometry experiments \jonathan{are combined} with dynamic micromagnetic simulations to connect microstructural features, magnetization dynamics, and macroscopic heat dissipation. Beyond macroscopic heating metrics, the heat generation is resolved at the intra-particle level, uncovering a heterogeneous landscape of localized ``hot spots'' with nanometer spatial and nanosecond temporal resolution. The results demonstrate that grain size acts as a key experimentally tunable parameter, balancing anisotropy disorder and pinning strength, thereby controlling both the magnitude and spatio-temporal distribution of heat release \jonathan{within the particle}. In particular, nanoflower architectures composed by larger grains deliver larger heat generation, while the smaller grains offer a deeper intra-particle pinning landscape, which effectively redistributes the heat generation over extended time windows. Together, our results provide a mechanistic framework linking nanoparticle microstructure to magnetic heating and establish design principles for optimizing nanoflowers as magnetic hyperthermia transducers.

\end{abstract} 


\maketitle


\section*{Introduction}
Iron oxide nanoparticles (IONPs) represent a class of materials with excellent magnetic properties and \jonathan{with} suitable biocompatibility for clinical use \cite{wu2024roadmap, pankhurst2009progress, pankhurst2003applications, colombo2012biological}. One of the distinctive properties of IONPs is their capability to generate heat under the application of an alternating (AC) magnetic field. Acting as nanotransducers converting magnetic field energy into heat, this capability of the IONPs is primarily leveraged in magnetic hyperthermia therapy (MHT) to disrupt the activity of cancerous cells \cite{diaz2025preclinical, perigo2015fundamentals}. In recent years, its scope has been expanded to integrate the heat in magnetic particle imaging \cite{rivera2021emerging}, neuromodulation \cite{gavilan2025magnetic}, drug delivery applications \cite{kumar2011magnetic}, as well as emerging fields such as catalysis and environmental remediation~\cite{gallo2024magnetic, gavilan2025magnetic}. \newline

The heating capability of IONPs has been extensively demonstrated experimentally \cite{jefremovas2021nanoflowers, bender2018relating, mekseriwattana2025shape, diaz2025preclinical}. This includes early evidence of heat generation at subnanometer distances from the particle surface \cite{riedinger2013subnanometer}, and cellular studies showing that magnetic hyperthermia–induced damage is not associated with a global temperature rise but rather with localized, single-cell mechanisms \cite{blanco2016real}. However, the microscopic origin of this heat generation remains largely unexplored. In particular, little is known about how heat is produced and distributed within individual nanoparticles, especially beyond the single-domain limit, where spin textures directly affecting the macroscopic magnetic properties emerge~\cite{lappas2019vacancy, prajapati2024transforming, vivas2020toward} and magnetic disorder becomes prominent~\cite{jefremovas2026coercivity}. Elucidating the spatial and temporal intra-particle heat landscape, \textit{i.e.}, the nanoscale distribution of ``hot spots'', and linking it to structural and magnetic inhomogeneities is therefore essential to rationally tailor heat generation and enable optimized nanoparticle design for magnetic hyperthermia. This is particularly relevant in multi-core ensembles, where their multi-grain structure affect markedly their macroscopic magnetic properties even within the nominal single-domain regime \cite{bender2018relating, rizzo2026gram}. \newline

\begin{figure*}
\centering
\resizebox{2.0\columnwidth}{!}{\includegraphics{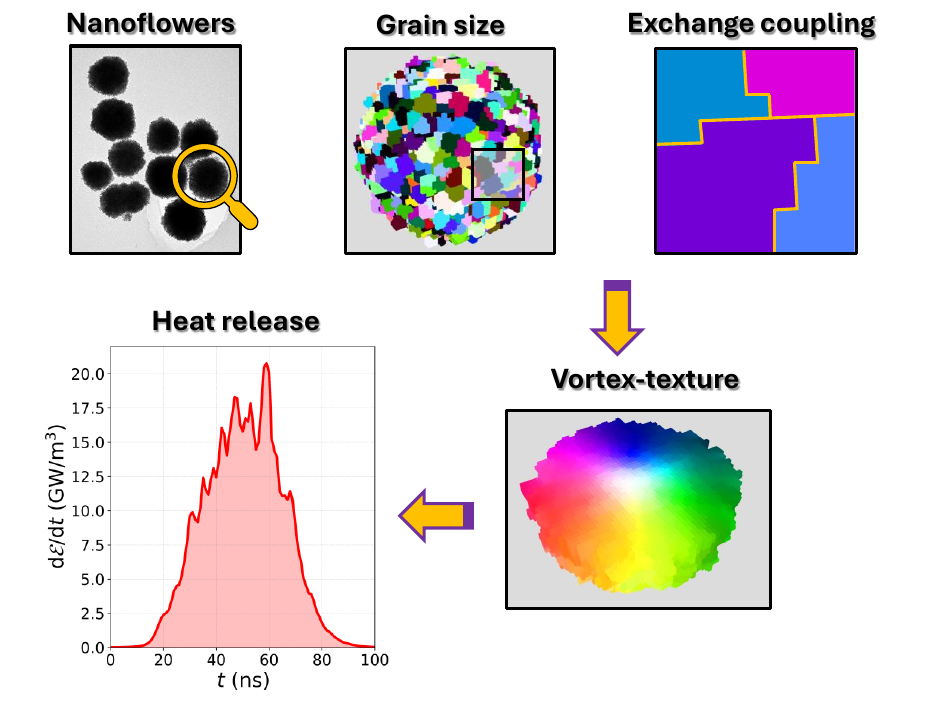}}
    \caption{Iron oxide nanoflowers are multicore aggregates of nanocrystals (commonly referred to as \textit{grains}) of tailorable size with promising prospects in MHT. Our model assigns each grain a random uniaxial anisotropy axis, $\vec{K_u}$, represented by a different colour in the figure. The boundaries in between those grains, highlighted in yellow, are a source of structural and magnetic disorder. We account for them through a reduction of the exchange-coupling at inter-grain boundaries, which together with the grain anisotropy, define the disorder landscape of the NFs. For maghemite NFs above the single-domain threshold ($d \gtrsim 70~\mathrm{nm}$ \cite{jefremovas2026coercivity, di2012generalization}), the magnetic texture folds into a vortex configuration, formed by a vortex core (represented in white) and a perpendicular flux-closure texture (each magnetization direction represented in a different color). In this work, we resolve the heat release under MHT conditions at local scale, with nanometer spatial and nanosecond temporal resolution.}
    \label{centrar_el_tiro}
\end{figure*}

Using a micromagnetic framework, we describe the magnetization dynamics of multicore iron-oxide nanoparticles, referred to as nanoflowers (NFs) (see Figure~\ref{centrar_el_tiro}), whose magnetization folds into a vortex for sizes $d> 70~\mathrm{nm}$. Such a spin texture is highly relevant for biomedical applications, as it encompasses a vortex core, carrying a net magnetization which can contribute to heating or serve as a vector for drug delivery~\cite{ho2011monodisperse, goiriena2016high, goiriena2020disk, usov2018magnetic, gavilan2021magnetic, jefremovas2021nanoflowers, hugounenq2012iron}, surrounded by a \jonathan{whirling} flux-closure \jonathan{domain}, which minimizes stray magnetic fields, thereby preventing undesired magnetic aggregation~\cite{serantes2021nanoparticle, etheridge2014accounting}. Particularly, the morphology of NFs offers a defect-rich, structurally tailorable platform via the grain size~\cite{gavilan2017formation, gallo2024magnetic}, where the influence of intra-particle defects into the heating response can be determined.\newline

In this work, we capture the heat generation of several NFs of different particle and grain size, evaluating their heating performance measured under AC fields within the range of MHT clinical conditions. Guided by these experiments, we reproduce the realistic morphology of NFs and investigate their dynamics within the micromagnetic framework. This allows us to link their structural features to the dynamics of the vortex texture and its related macroscopic heat generation.  Our simulations elucidate the functional role of spin disorder in the macroscopic heating metrics and resolves the heat generation within individual particles, revealing a heterogeneous landscape of localized ``hot spots'' quantified at nanometer spatial and nanosecond temporal resolution. Together, our results elucidate the functional role of spin disorder in magnetic heating and establish a direct connection between nanoparticle microstructure, micromagnetic texture, and heat dissipation, providing a mechanistic basis for the rational optimization of magnetic hyperthermia. \newline 

\section*{Materials and methods}

\subsection{Experimental methods}

Maghemite nanoflowers with the smaller grains (7 nm and 10 nm) were synthesized following a scaled-up version of our previously reported modified solvothermal polyol protocol \cite{gavilan2017formation}. Briefly, 4.37 g of iron(III) chloride hexahydrate (FeCl$_3\cdot$6H$_2$O, $\geq$98\%, Sigma--Aldrich) were dissolved in 870 mL of ethylene glycol under magnetic stirring at 100~$^\circ$C for 1 h. Subsequently, 28.4 g of sodium acetate and poly(vinylpyrrolidone) (PVP, $M_\mathrm{w} \approx 40\,000$, Sigma--Aldrich) were added to the solution. The amount of PVP was varied to tune the particle size: 90 g were used to obtain nanoflowers with an average diameter of $\approx 170$ nm, whereas 45 g were used to produce larger particles ($d \approx 220$ nm). The mixture was stirred for $\approx 30$ min until complete dissolution of the reagents, yielding a homogeneous precursor solution. The solution was then transferred to a Teflon-lined stainless-steel autoclave (1000 mL capacity) and heated at 200~$^\circ$C for 16 h. After naturally cooling to room temperature, the resulting particles were magnetically separated and washed several times with ethanol.\newline

The obtained particles were subsequently subjected to an acidic oxidation treatment. First, the particles were washed with 300 mL of nitric acid (HNO$_3$, 65\%; Sigma--Aldrich) solution (2 M) and magnetically separated. They were then redispersed in 200 mL of iron(III) nitrate nonahydrate (Fe(NO$_3$)$_3\cdot$9H$_2$O, $\geq$99.95\%; Sigma--Aldrich) solution (0.6 M) and heated at 90~$^\circ$C for 1 h. After cooling to room temperature, the particles were magnetically separated again, washed with 300 mL of HNO$_3$ (2 M), rinsed three times with an acetone/water mixture (30:70 v/v), and finally dried at 50~$^\circ$C overnight.\newline

Nanoflowers with larger grains (21nm and 25 nm) were synthesized under analogous solvothermal conditions. For particles with an average diameter of $d = 170$ nm, 10.4 g of FeCl$_3\cdot$6H$_2$O and 3.2 g of trisodium citrate dihydrate (99\%; Sigma Life Siences) were dissolved in 320 mL of ethylene glycol ($\geq$99\%, Sigma--Aldrich), followed by the addition of 20 g of sodium acetate trihydrate ($\geq$99\%, Sigma--Aldrich) under magnetic stirring. For larger particles ($d \approx 250$ nm), 21.62 g of FeCl$_3\cdot$6H$_2$O and 6 g of trisodium citrate were dissolved in 376 mL of ethylene glycol, after which 40 g of sodium acetate trihydrate were added. In both cases, the mixtures were vigorously stirred for 30 min, transferred into a Teflon-lined aluminium autoclave (1000 mL), and heated at 200~$^\circ$C for 16 h. After natural cooling to room temperature, the resulting black precipitate was magnetically separated and washed three times with distilled water by magnetic decantation. The final product was subsequently subjected to the acidic oxidation treatment described above and dried at 50~$^\circ$C overnight.\newline

Transmission electron microscopy (TEM) was performed to determine the morphology and particle-size distribution of the NFs. X-ray diffraction (XRD) was performed to identify the phase and nanocrystallite (grain) size of the NFs. Further details are included in Supplemental Material S1.\newline

\textcolor{black}{AC magnetometry was performed using a commercial AC magnetometer system (``AC Hyster'', Nanotech Solutions, Spain). The system is designed for characterising liquid nanoparticle samples of 40 $\mu$L. Hysteresis loops have been measured for the 4 different NF samples of several diameter and grain sizes: $d = 170~\mathrm{nm}$, with grain sizes of $g_{s} = 21~\mathrm{and}~7~\mathrm{nm}$, and $d = 220~\mathrm{and}~250~\mathrm{nm}$, with $g_{s} = 10~\mathrm{and}~25~\mathrm{nm}$, respectively, at three different frequencies, $f$ = 10, 25, and 100 kHz, with AC field amplitudes from $\mu_{0}h_{\mathrm{AC}}$ = 3.83 kA/m to 25.81 kA/m~\textit{i.e.}, all within the clinical safety limits $h\cdot f \leq 5~\mathrm{GA/ms}$~\cite{hergt2007magnetic, hergt2009validity}}. \newline
\subsection{Numerical methods}
\subsection{Micromagnetic description of maghemite nanoflowers}
\jonathan{Micromagnetic simulations were performed using Mumax3\cite{Vansteenkiste2014}. The NFs were generated starting from a cubic simulation region to which a Voronoi tessellation was applied to divide the volume into discrete regions corresponding to individual material grains \cite{Lel2014}. A perfectly spherical geometry with an initial diameter $d$ was then overlaid on this tessellated cube. Only the grains whose Voronoi centers were located inside the sphere were retained as part of the particle geometry, while grains with centers outside the sphere were removed. This procedure yields a nanoflower-shaped particle composed of multiple irregular grains (see Figure~\ref{centrar_el_tiro}). Intra-particle disorder was included by assigning a random uniaxial anisotropy direction $K_{u}$ to each grain and by modifying the inter-grain exchange coupling at grain boundaries. Additional details on the simulations are included in the Supplemental Material and Fig. S4 therein.}\newline
Critically, we identify both parameters to be controlled by an experimentally-tailorable parameter, the grain size~\cite{gavilan2017formation, gallo2024magnetic}, which effectively determines (i) the intra-particle regions sharing the same $K_{u}$; and (ii) the amount of intra-particle grain boundaries, which determines the pinning landscape. The intra-particle pinning was captured by modelling the magnetic coupling between adjacent grains via rescaling the exchange parameter $A$ at the grain boundaries by a factor $k$ lying between 0 and 1. Material parameters typical for iron oxide were used, including saturation magnetization $M_{s} = 400~\mathrm{kA/m}$~\cite{roca2007effect, shokrollahi2017review}, exchange stiffness $A=10~\mathrm{pJ/m}$ \cite{sinaga2024neutron}, uniaxial magnetocrystalline anisotropy $K_{u} = 10~\mathrm{kJ/m^{3}}$ \cite{gross2021magnetic, borchers2025magnetic, pisane2017unusual, roca2007effect} and Gilbert damping parameter $\alpha = 0.1$ \cite{allia2011dynamic, ccam2026tuning}. A time-dependent magnetic field $h(t) = h_{\mathrm{AC}} \cos\!\left(2\pi f\, t\right)$ is applied along the $z$-axis direction, with $f = 300\;\mathrm{kHz}$, to calculate the heat losses per cycle. Using Mumax3, we numerically integrate the Landau-Lifshitz-Gilbert (LLG) equation:

\begin{equation}
\frac{d\mathbf{m}}{dt}
= -\frac{\gamma}{1+\alpha^2}\left[
\mathbf{m}\times\mathbf{B}_{\mathrm{eff}}
+ \alpha\,\mathbf{m}\times\left(\mathbf{m}\times\mathbf{B}_{\mathrm{eff}}\right)
\right],
\label{LLG}
\end{equation}

and link the torque exerted on the magnetization to the magnetic energy dissipated via Eq.~\ref{calorina}\cite{munoz2020disentangling}: \newline
\begin{equation}
    \frac{d\mathcal{E}}{dt}
= \frac{\alpha \gamma M_s}{1+\alpha^2}\, (\mathbf{m}\times\mathbf{B}_{\mathrm{eff}})^2
\label{calorina}
\end{equation}

To date, this expression has only been applied to nanoparticles within the macrospin approximation~\cite{leliaert2021individual, ortega2023estimating}. Our work moves a step forward by applying it to \jonathan{the individual finite-difference cells in the simulation to investigate the heat dissipation in nanoparticles with internal magnetization inhomogeneities}, for which large-scale simulations are performed. This dissipated energy can be transformed into the specific loss power or, equivalently, specific absorption rate, $SAR$\footnote{Although a distinction between $SLP$ (intrinsic heat generation) and $SAR$ (experimental quantification of heat generation) is  sometimes distinguished on conceptual grounds, both metrics are conceptually equivalent and quantify the same underlying heating process. Their formulation differs only in their normalization factor, with $SAR$ defined per mass concentration in the medium ($c$) and $SLP$ per intrinsic material density ($\rho$).} parameter quantifying the heating performance from the hysteresis losses, via:
\begin{equation}
SLP = \frac{f}{\rho} \mu_0 \oint M(H)\,\mathrm{d}H ,
\label{SLP_equ}\end{equation}

and thus:

\begin{equation}
SLP = \frac{f}{\rho} \mu_0 \oint  M(H)\,\mathrm{d}H
=
\frac{f}{\rho}\mathcal{E}_{\mathrm{cycle}}^{(V)}
\end{equation}

Being $\rho$ the density of the material, which for the present case of maghemite, is $\rho = 4.9~\mathrm{g/cm^{3}}$ \cite{guivar2014structural}. \newline

\section*{Results and discussion}


To link microstructural disorder to heat generation under magnetic hyperthermia conditions, we identify the grain size, $g_{s}$, as the \jonathan{most relevant} physical control parameter. From a micromagnetic perspective, this parameter determines the amount of constituent grains, $N$, balancing thus both the role of inter-grain pinning boundaries and anisotropy disorder into the magnetization dynamics and heat generation. From an experimental perspective, the grain size can be precisely controlled via growth conditions at fixed nanoflower size \cite{gavilán2017formation, gavilan2021size, gavilan2021magnetic}, thereby enhancing the applicability of our conclusions. Our study combining experimental and simulation results provides unprecedented insights to the intra-particle mechanisms governing the heat generation, capturing the physics underneath the experimental observations. \newline
\section*{Grain size impact on heat generation: experimental evidence}

AC hysteresis loops measurements were performed in four different ensembles of nanoflowers with diameter $d$ and grain size $g_{s}$: $(d, g_{s}) = (170, 21), (170, 7), (220, 10)~\mathrm{and}~(250, 25)~\mathrm{nm}$, from which the Specific Absorption Rate ($SAR$) was extracted from the associated hysteresis losses (see Supplemental Material Figure S2) via:

\begin{equation}
    SAR = \frac{f}{c}\cdot A = \frac{f}{c} \cdot \mu_{0} \oint  M_{t} dH_{t}
\end{equation}

where $M_{t}$ is the instantaneous magnetization at time $t$, $H_{t}$ denotes the sinusoidal magnetic field of frequency $f$ at time $t$, and $c$ is the magnetic material weight concentration in the dispersing medium. These four ensembles of NFs constitute two representative pairs of NFs with similar diameter and different grain size, all being within the same magnetic regime, where the magnetization reversal is dominated by the flux-closure moments \cite{jefremovas2026coercivity}. \newline

\begin{figure*}
\centering
\resizebox{2.0\columnwidth}{!}{\includegraphics{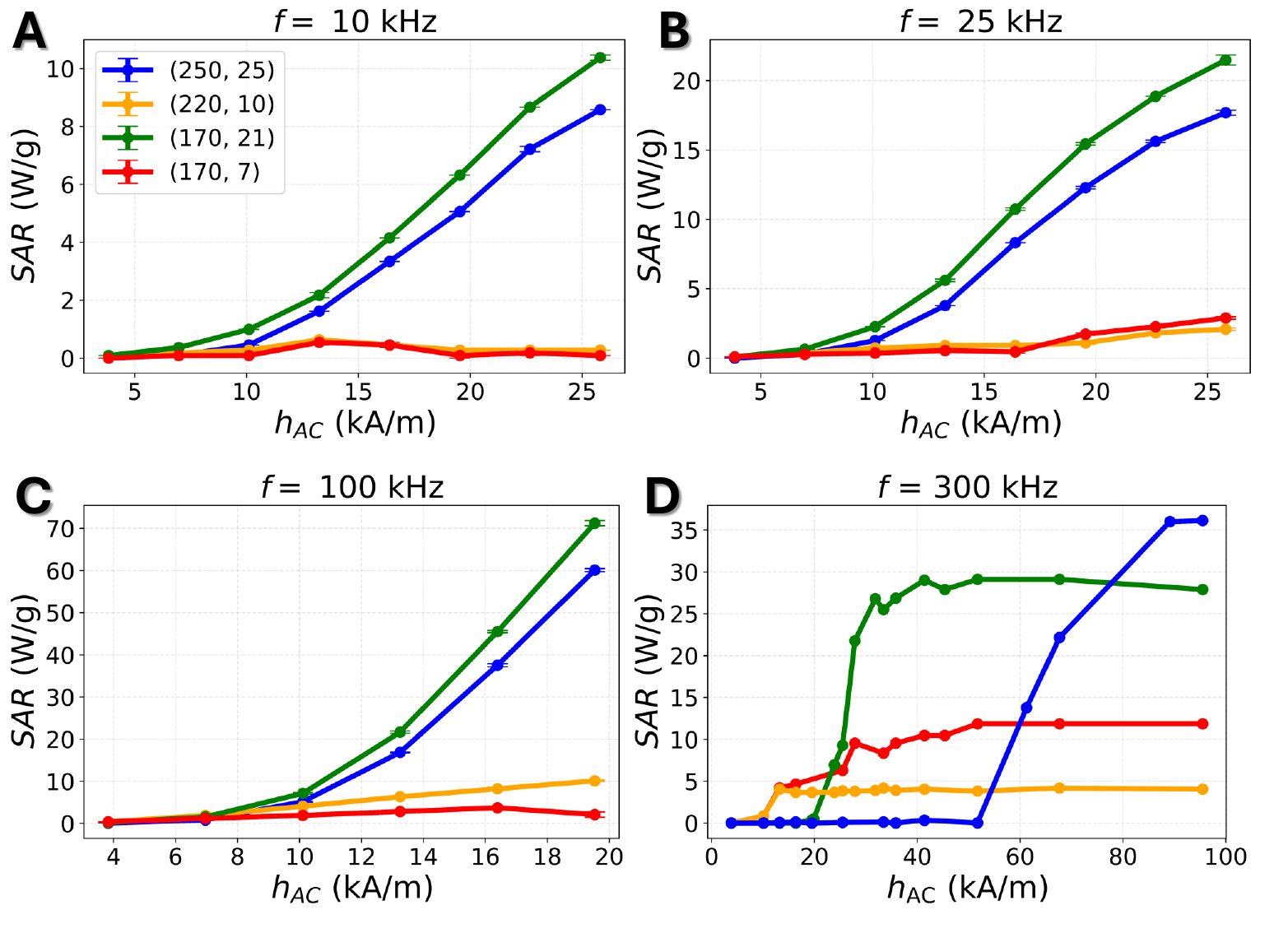}}
    \caption{Specific absorption rate, $SAR$, \textit{vs.} field intensity $h_{\mathrm{AC}}$, for the 4 NF ensembles measured at \textbf{a)} $f = $ 10, \textbf{b)} $f = $ 25, and \textbf{c)} $f = $ 100 kHz. In all cases, the NF with the smaller grain sizes $g_{s}$ display almost negligible $SAR$ throughout the explored field range. \textbf{d)} Numerical results of the corresponding $(d, gs)$ realizations at $f = 300~\mathrm{kHz}$. The simulations are consistent with the experiments: NFs with larger grains generate more heat than their smaller grain counterparts.}
    \label{SAR}
\end{figure*}

Figure~\ref{SAR}\textbf{~(A)-(C)} displays the $SAR$ as a function of the field amplitude $h_{\mathrm{AC}}$ for the three measured excitation frequencies $f$. For the samples with larger grain size, a clear $SAR$ signal develops for $h_{\mathrm{AC}} \gtrsim 10.11~\mathrm{kA/m}$, increasing nearly linearly with $h_{\mathrm{AC}}$ and $f$ over a broad range. For $f = 10$ and $25~\mathrm{kHz}$, the $SAR(h_{\mathrm{AC}})$ curves start to bend above $h_{\mathrm{AC}} \gtrsim 19.53~\mathrm{kA/m}$, indicating a progressive approach to $SAR$ saturation. In contrast, at the highest frequency, $f = 100~\mathrm{kHz}$, the response remains far from saturation within the explored field window, implying substantial headroom for further improving performance at higher field amplitudes and frequencies. A linear extrapolation of the $f = 100.0~\mathrm{kHz}$ data beyond $h_{\mathrm{AC}} = 19.53~\mathrm{kA/m}$ suggests that $SAR \cong 200~\mathrm{W/g}$ could be reached at $h_{\mathrm{AC}} = 39.79~\mathrm{kA/m} \cong 50~\mathrm{mT}$ (see Supplemental Material Figure S3), a value comparable to that reported for commercial 40~nm NFs at $f = 130~\mathrm{kHz}$ in Ref.~\cite{jefremovas2021nanoflowers}.

This result is particularly encouraging from the perspective of clinical translation \cite{beola2020intracellular}. Above $d = 70~\mathrm{nm}$, the magnetization of the NFs is expected to fold into a vortex configuration~\cite{moya2024unveiling, jefremovas2026coercivity}. Such a magnetic texture strongly suppresses stray fields, significantly reducing the tendency of magnetic nanoparticles to agglomerate. Since particle agglomeration is known to critically deteriorate heating performance in \textit{in vivo} environments~\cite{etheridge2014accounting, guibert2015hyperthermia, ortega2023estimating, goiriena2020disk}, the fact that vortex-state NFs can sustain heating efficiencies comparable to those of smaller single-domain particles represents a highly desirable scenario: efficient heat generation combined with potentially improved colloidal stability under physiological conditions.\newline

On the other hand, samples with smaller grain sizes exhibit almost negligible $SAR$ at low $f$. Only for $f = 100~\mathrm{kHz}$ does the 220~nm sample with $g_{s} = 10~\mathrm{nm}$ develop a measurable $SAR$ signal above $h_{\mathrm{AC}} = 10.11~\mathrm{kA/m}$, which nevertheless remains nearly seven times smaller than the values obtained for the (250, 25)~nm or (170, 21)~nm NFs. This behaviour suggests that, for small grain sizes, the intra-particle magnetic configuration is characterized by stronger pinning, effectively leading to larger energy barriers for magnetization reconfiguration. This interpretation is consistent with the observation that larger $f$ and $h_{AC}$ are required to overcome these barriers and trigger a measurable heat release in NFs with smaller grains. Micromagnetic simulations will further shed light on the microscopic origin of this behaviour.

To elucidate these experimental observations, we performed micromagnetic simulations mimicking the experimental conditions by calculating the heat release per cycle over magnetic fields ranging from $h_{\mathrm{AC}} = 3.8~\mathrm{kA/m}$ to $h_{\mathrm{AC}} = 51.7~\mathrm{kA/m}$ at $f = 300~\mathrm{kHz}$. Figure~\ref{SAR}\textbf{(D)} includes our numerical results for 170~nm NFs with 21 and 7~nm grain sizes, together with $(d, g_{s}) = (220, 10)~\mathrm{and}~(250,25)~\mathrm{nm}$. The calculated $SAR$ values reproduce the experimental trends: NFs composed of larger grains, and thus fewer grain boundaries, exhibit enhanced heat dissipation compared to their small-grain counterparts at the probed fields. Notably, the saturated $SAR$ is approximately three times larger for $g_{s} = 21~\mathrm{nm}$ than for $g_{s} = 7~\mathrm{nm}$, increased to almost 7 times for the 220-250 pairs. 


Furthermore, the comparison between experiments and simulations provided in Figure~\ref{SAR} underpins the role of thermal fluctuations. Simulations for 170~nm NFs with 21~nm grains show a threshold $h_{\mathrm{AC}} \approx 24~\mathrm{kA/m}$ for heat release, approximately twice the experimental value, where $SAR \geq 7~\mathrm{W/g}$ is already observed at $h_{\mathrm{AC}}\approx 12.5~\mathrm{kA/m}$. The same trend follows for the larger NFs, where a threshold value of $h_{\mathrm{AC}} \approx 60~\mathrm{kA/m}$ for the 250~nm-sized NFs, above the experimental value of $h_{\mathrm{AC}}\approx 10~\mathrm{kA/m}$ detected, whereas a reduced one of $h_{\mathrm{AC}} \approx 12~\mathrm{kA/m}$ for the 220~nm-ones is observed. We ascribe the discrepancy between experimental and numerical results to thermal fluctuations. Indeed, although thermal fluctuations are insufficient to dominate magnetization reversal of such large nanoparticles, they effectively lower energy barriers. When the characteristic barrier-crossing time becomes comparable to the AC-field period, thermally assisted switching becomes allowed, giving rise to non-zero energy dissipation. \newline

Our numerical results underscore that the poorer heating metrics of the NFs with smaller grains should be ascribed to intrinsic magnetic arguments, rather than to thermal fluctuations. Both the experiments and simulations consistently showcase a reduced heat generation for NFs with smaller grains. Indeed, the increased density of intra-particle boundaries further elevates the energy barriers, imposing larger field excitations (both amplitude and frequency wise) to overcome intra-particle boundaries and trigger a significant heat release. The increased density of intra-particle boundaries also plays a role in the larger field excitation needed to trigger a $SAR$ response in the larger NFs observed in our simulations. The $(d, g_s) = (250,25)~\mathrm{nm}$ NFs double the amount of constituent grains of the $(d, g_s) = (170, 21)~\mathrm{nm}$ ones, requiring thus larger field excitation for the onset of the $SAR$, achieving also larger $SAR$ at saturation. Indeed, thermal fluctuations, present in the experiments, may assist the switching, lowering the field excitation requirements and, consequently, the threshold for these larger NFs. \newline

Finally, size polydispersity and particle agglomeration may further contribute to discrepancies between experiments and simulations. While our calculations consider heat dissipation from a single NF, experiments probe ensembles at Fe concentrations of $\sim 2$--$4~\mathrm{mg/cm^{3}}$, corresponding to particle concentrations $\sim 10^{11}~\mathrm{NFs/cm^{3}}$. Ensemble effects, including size distributions \cite{jeun2009effects,khandhar2011monodispersed} and interparticle interactions arising from agglomeration \cite{etheridge2014accounting, guibert2015hyperthermia, ortega2023estimating}, are therefore expected to influence the measured heating metrics. \newline

All in all, our results elucidate the key influence of grain size into the heating power of NFs. Nanoparticles with smaller grains provide a small heat generation at the probed fields, whereas their larger grain size counterparts deliver heat comparable to smaller commercial counterparts~\cite{jefremovas2021nanoflowers}, with the key advantage of being less prompt to agglomeration. In the next sections, we will use micromagnetic simulations to capture the underlying fine balance between inter-grain pinning and anisotropy landscape in the heat release, observing a predominant role of the individual grain anisotropy over the inter-grain pinning boundaries. \newline


\section*{Grain size impact on the heat generation: Numerical study}

Our experimental results reveal a key influence of the grain size into the heat dissipation, with a privileged role of larger grains compared to the smaller ones. These results are reproduced by our numerical simulations, indicating the microscopic origin of such effect. We herein build further on the numerical side to capture the intra-particle mechanisms giving rise to such differences.\newline 

To unambiguously evaluate the influence of $g_{s}$ on heat generation, we compare NFs with $d = 170$ nm, whose magnetization dynamics is dominated by flux-closure moments~\cite{jefremovas2026coercivity}, with smaller NFs, $d = 100~\mathrm{nm}$, where the dynamics is instead governed by the magnetic moments within the vortex core~\cite{jefremovas2026coercivity}. This comparison allows us to extend the conclusions on grain-size effects to the whole vortex regime, beyond the specific flux-closure ensembles investigated experimentally in this work. For completeness, we have included in Supplemental Material Section S5 the cases of $d = 220$ nm and $d = 250$ nm, which show qualitatively the same behaviour as $d = 170$ nm. \newline

\begin{figure*}
\centering
\resizebox{1.7\columnwidth}{!}{\includegraphics{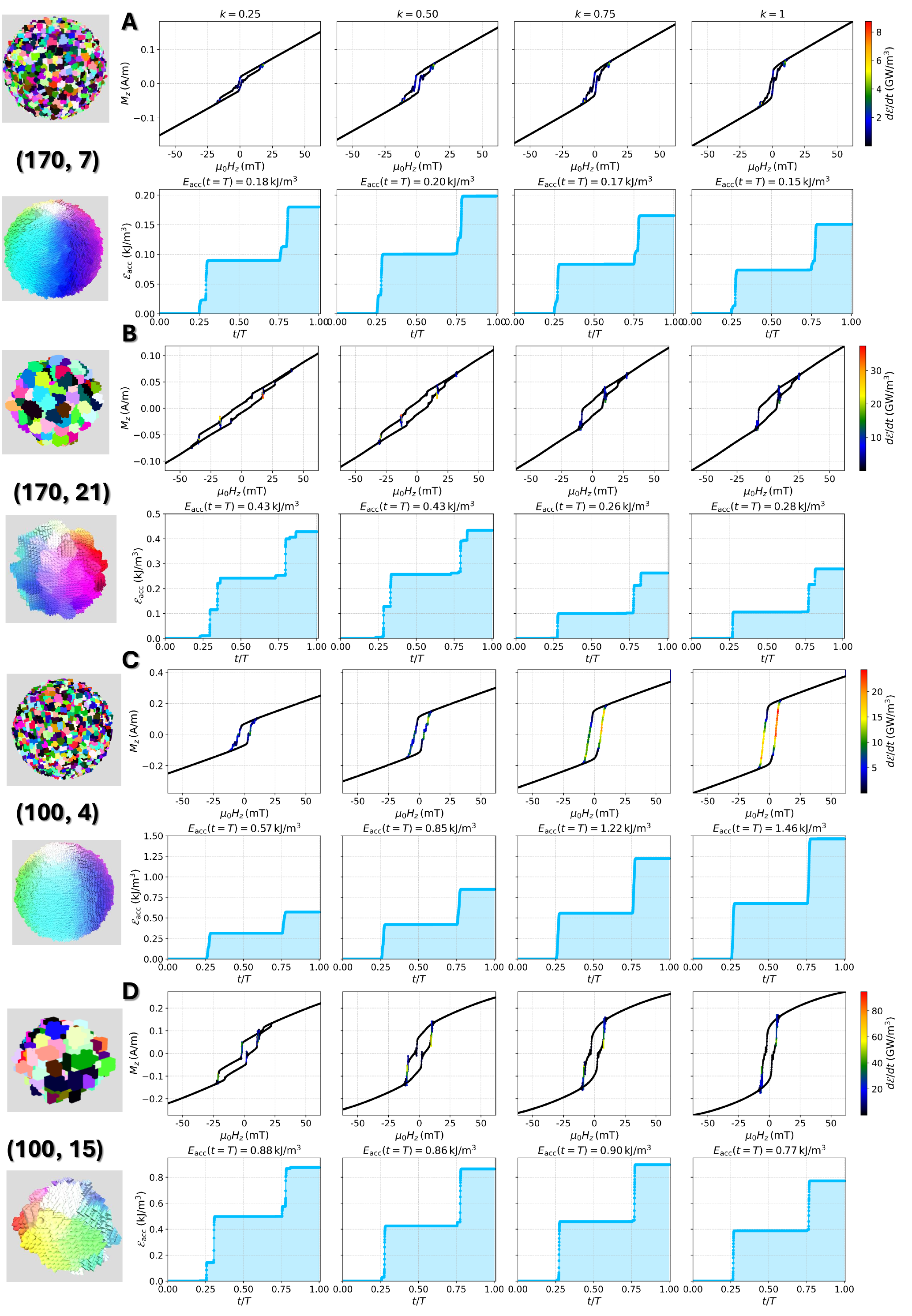}}
    \caption{Magnetization as a function of applied field (top row) and released energy $\mathcal{E}(t)$ (bottom row) over one AC-field period $T = 1/300~\mathrm{ms}$ for NFs with diameters of $d=170~\mathrm{nm}$ and $g_{s}=21~\mathrm{nm}$ (\textbf{A}) and $7~\mathrm{nm}$ (\textbf{B}), and $d=100~\mathrm{nm}$ and $g_{s}=4~\mathrm{nm}$ (\textbf{C}) and $15~\mathrm{nm}$ (\textbf{D}). The magnetization traces are colour-coded by the instantaneous heat generation, with red highlighting the points of maximum dissipation, coinciding with the jumps observed in the magnetization consequence of the reversal, \textit{i.e.} moments of maximized torque. The step-like increases in $\mathcal{E}(t)$ coincide with these dissipation bursts. Sketches represent in top the grain anisotropy, where each colour represents a $\vec{K_{u}}$ direction, and the vortex configuration (bottom), where each colour represents a direction of the magnetization.}
    \label{grain_size}
\end{figure*}

Figure~\ref{grain_size} includes the results corresponding to $d = 170~\mathrm{nm}$ with $g_{s} = 7$ and $21~\mathrm{nm}$, and $d = 100~\mathrm{nm}$ with grain sizes $g_{s} = 4$ and $15~\mathrm{nm}$. These provide pairs of realizations with comparable numbers of constituent grains $N$. Specifically, the configurations $(d,gs) = (100~\mathrm{nm},4~\mathrm{nm})$ and $(170~\mathrm{nm},7~\mathrm{nm})$ comprise similar grain counts, with $N \simeq 8181$ and $N \simeq 7500$, respectively, while the configurations $(100~\mathrm{nm},15~\mathrm{nm})$ and $(170~\mathrm{nm},21~\mathrm{nm})$ yield $N \simeq 160$ and $N \simeq 278$. The use of comparable grain numbers $N$ across these size regimes enables to extract and identify the different functional role of intra-particle disorder within the two magnetic dynamics regimes. \newline

Figure~\ref{grain_size} includes the $M_{z}$ \textit{vs.} $H_{z}$ graphs, together with the corresponding magnetic energy accumulated per AC cycle, $\mathcal{E}_{\mathrm{acc}}$. As it can be seen in the colour-coded magnetization, realizations with smaller $g_{s}$ showcase a sustained, progressive heat release, over the whole reversal process (see top rows in \textbf{(A)} and \textbf{(C)}), whereas for the larger $g_{s}$, the heat release is concentrated at the extremes of the reversal process (see \textbf{(B)} and \textbf{(D)}), regardless of the vortex core- or flux-closure-dominated regimes. We ascribe the vortex-core pinning at the grains (nanocrystallites) as the main source generating a torque for magnetization reversal, which translates into heat. Smaller grain sizes $g_{s}$ imply a larger number of constituent grains $N$, which generates a denser and more complex pinning landscape, characterized by a higher density of inter-grain boundaries and a broader distribution of local anisotropy axes, which in turn, enhance the coherence and rigidity of the vortex core. The field-driven motion of a more rigid vortex core through this disordered landscape requires thus increased work against pinning forces, resulting in enhanced instantaneous energy dissipation where the energy generation is determined by the depinning and motion of the vortex core. On the other side, the smaller amount of $N$ of the larger grains result in a vortex core comparatively fragmented or strongly distorted, and the reversal proceeds through more local, incoherent rotations, which reduces the net torque and the associated energy dissipation. \newline

The total energy released per cycle, $\mathcal{E}_{\mathrm{acc}}$, points consistently to vortex core-dominated realizations to deliver larger values (see bottom rows in \textbf{(C)} and \textbf{(D)}) compared to the flux-closure ones (\textbf{(A)} and \textbf{(B)}). The fact that the magnetization reversal of flux-closure-dominated ensembles occurs partially perpendicular to the $z$--direction, whereas the vortex core-dominated follow a reversal path purely along the $z$--axis~\cite{jefremovas2026coercivity} smoothens the torque exerted by the driving field into the magnetization, resulting in a reduced heat generation for the larger flux-closure-dominated NFs. The different magnetization-reversal mechanism (forcing the vortex core to lay perpendicular to the $z$--axis, or keeping it purely along the $z$-axis) affects the $\mathcal{E}_{\mathrm{acc}}$ and its dependence on the inter-grain coupling $k$. Within the vortex-regime, we detect a systematic increase with $k$ for $(d,gs) = (100~\mathrm{nm},4~\mathrm{nm})$ (see Figure~\ref{grain_size} \textbf{(A)}). In this case, a larger inter-grain exchange coupling yields a more robust and coherent vortex-core profile, which in turn results in an enhanced torque and related energy release. On the contrary, within the flux-closure-dominated regime, the collective dipolar-mediated rotation smoothens the dynamics, weakening the dependence on $k$. \newline



\section*{Nanoscale mapping of intra-particle heat generation}

Up to this point, we have examined the functional role of intra-particle magnetic disorder into macroscopic parameters, magnetization $M_{z}$ and magnetic energy accumulated per AC field cycle, $\mathcal{E}_{\mathrm{acc}}$. We herein resolve the heat generation microscopically, identifying the presence of localized ``hot spots'' at intra-particle level, connecting them to the magnetization reversal stage at nanosecond time resolution.\newline

\begin{figure*}
\centering
\resizebox{1.8\columnwidth}{!}{\includegraphics{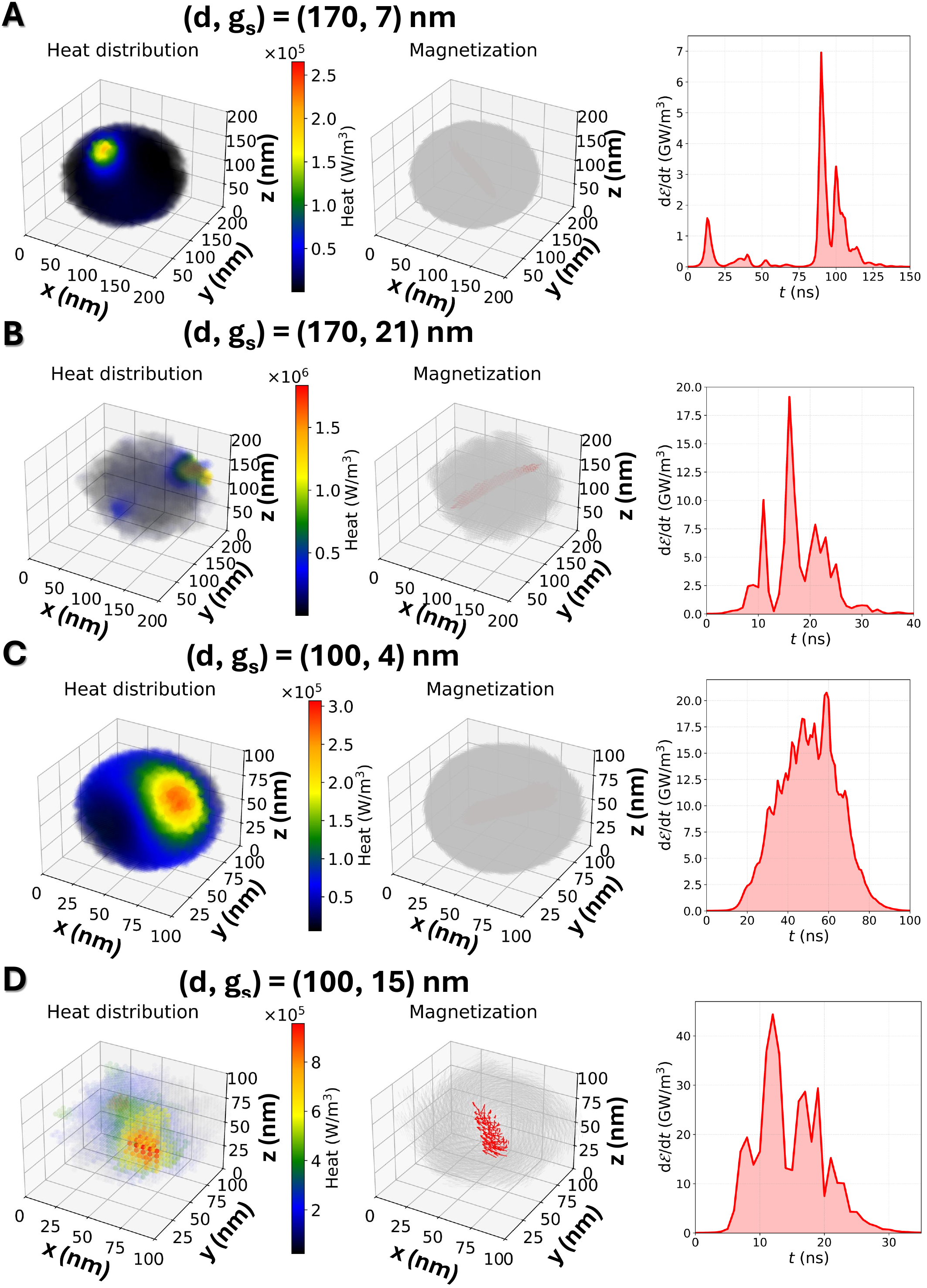}}
    \caption{3D mappings of heat generation together with the corresponding 3D magnetization configurations and instantaneous heat release $d\mathcal{E}/dt$ as a function of time for $(d, g_{s}) = (170, 7), (170, 21), (100, 4)$ and $(100, 15)~\mathrm{nm}$ (\textbf{(A)}–\textbf{(D)}, respectively). In all cases the intergrain exchange coupling parameter is fixed to $k = 1$, and the external alternating magnetic field is applied at an amplitude of $\mu_{0}h = 62~\mathrm{mT}$ and frequency $f = 300~\mathrm{kHz}$. The colour bar indicates the intensity of the heat release, ranging from black (minimum) to red (maximum). Magnetic moments within the vortex core are shown as red arrows, while flux-closure moments are represented in grey. }
    \label{instantaneous_heat}
\end{figure*}

Figure~\ref{instantaneous_heat} showcases the spatial distribution of the local heat release at nanometer resolution, at the instant of maximum dissipation for each configuration. In all cases, this moment corresponds to a vortex configuration where the core is oriented perpendicular to the applied field along the $z$ axis (see 3D magnetization plots in the middle panel of Figure~\ref{instantaneous_heat}), maximizing the instantaneous misalignment between the magnetization and $\mathbf{B}_{\mathrm{eff}}$ and thus the heat generation. For the NFs with smaller grains, and thus a larger number of constituents, the heat release is spatially distributed over multiple regions, reflecting a complex pinning landscape where the vortex core interacts with numerous grain boundaries and local anisotropy directions. This results in an extended dissipation pattern, consistent with a gradual and sustained reversal process. In contrast, NFs with larger grains exhibit highly localized and less dense hot spots, indicating a rapid and collective reorientation of the magnetic texture. In this regime, the reduced pinning landscape allows the vortex state to collapse or rotate abruptly once the Zeeman term becomes the dominant factor within the effective field $B_\mathrm{{eff}}$, concentrating the dissipation both spatially and temporally. \newline

Figure~\ref{instantaneous_heat} further includes the instantaneous heat release $d\mathcal{E}/dt$ vs. $t$, along with the visualizations of the heat and magnetization resolved at nanometer resolution (cell size) for $d = 170~\mathrm{and}~100~\mathrm{nm}$ with $g_{s} = 7,~21,~\mathrm{and}~4,~15~\mathrm{nm}$, respectively. As it can be observed, the heat release for the smaller grains is sustained over the whole reversal process, which we identified to happen at $t\sim 100~\mathrm{ns}$. The heat release happens for the larger grains however a factor 4 faster, $t\sim 25~\mathrm{ns}$, concentrated at the beginning of the reversal process. Such distinct temporal profiles of the heat release can traced back to the reversal dynamics through Eq.~\ref{calorina}, which relates the instantaneous dissipation rate $d\mathcal{E}/dt$ to the  $(\mathbf{m}\times\mathbf{B}_{\mathrm{eff}})^2$ term. This term reflects the phase lag between the particle magnetization and the $B_\mathrm{{eff}}$ imposed by the driving AC field, from which the heating originates. In NFs with smaller grains, \textit{i.e.}, larger amount of intra-particle boundaries $N$, the dense and heterogeneous pinning landscape prevents rapid alignment of the magnetization with the effective field, sustaining moderate but nonzero values of $(\mathbf{m}\times\mathbf{B}_{\mathrm{eff}})^2$ over an extended time interval. As a consequence, the heat is released gradually throughout the reversal process, as indicated in the $M$ \textit{vs.} $B_{ext, z}$ plots included in Fig.~\ref{grain_size}. In contrast, NFs with larger grains offer a reduced pinning landscape, which reduces the rigidity of the vortex core. As a result, the strong initial misalignment of the vortex core with $\mathbf{B}_{\mathrm{eff}}$ can be easily corrected, delivering a large value of the torque $(\mathbf{m}\times\mathbf{B}_{\mathrm{eff}})$ over a short time interval. Following Eq.~\ref{calorina}, this leads to an intense but shorter dissipation of energy, followed by a negligible contribution once the magnetization realigns with the effective field. It is also worth mentioning the striking differences that intra-particle magnetic disorder introduces to the heat generation distribution, in both space and time domains. Figure S7 in Supplemental Material includes the calculated heat generation of a perfect $d = 50$ nm single-domain nanospheres. As observed, the heat release occurs at faster pace (within 5 nanoseconds) and spreads over the whole particle, which contrast to the case of the NFs, for which the grain anisotropy delays the heat release and the vortex magnetization concentrates the heat release at the vortex core.  \newline

The maximum value of the instantaneous heat release further reflects the interplay between reversal dynamics and internal microstructure. Among the systems studied, the largest peak values of $d\mathcal{E}/dt$ are observed for the $(100~\mathrm{nm},15~\mathrm{nm})$ and $(170~\mathrm{nm},21~\mathrm{nm})$ cases, while significantly lower maxima are found for the smallest grains $(100~\mathrm{nm},4~\mathrm{nm})$ and $(170~\mathrm{nm},7~\mathrm{nm})$, whose maxima are found at $\cong10.013~\mathrm{GW/m^{3}}$ and $\cong 7.821~\mathrm{GW/m^{3}}$, a factor 5 and 7 smaller, respectively. This trend highlights that the instantaneous heat release benefits from the abrupt reorganization of the magnetic texture during reversal, yet such abrupt changes are not beneficial for the time-integrated energy release, as showcased in Fig.~\ref{grain_size}. In fact, slower magnetization dynamics imposed by a richer pinning landscape lead to a sustained heat release distributed over time, rather than to short-lived, intense peaks. Such a temporally extended dissipation is advantageous not only to deliver a larger generated energy per AC cycle, yet it further avoids highly localized power spikes and promotes a more homogeneous and controllable thermal response, which is particularly desirable for magnetic hyperthermia applications~\cite{gavilan2021magnetic}.\newline

Our heating maps provide direct visual evidence that the internal granular structure governs not only the magnitude but also the spatial and temporal character of the heat release, linking microstructure to reversal dynamics and energy dissipation.\newline

\section*{Conclusions}

In this work, we link microstructural features to macroscopic heating performance by resolving the spatio-temporal magnetization dynamics of iron oxide NFs, thereby connecting intra-particle magnetic defects to heat generation at the particle scale. Starting from the experimental observation that NFs with larger grains deliver better heat release, we use a micromagnetic framework to capture the influence of intra-particle magnetic disorder into the macroscopic heat release, and to localize spatially and temporally the occurrence of ``hot spots'' at intra-particle level. \newline

By using the grain size as an experimentally controllable parameter, we have determined the crucial influence of structural disorder into the heat generation. Whereas the smaller grains offer a larger collection of inter-grain pinning boundaries, which in principle should on its own deliver a larger energy release, they broaden at the same time the anisotropy landscape, which facilitates core mobility through hopping between local anisotropy directions, highlighting the critical balance between both pinning and anisotropy landscapes. As a result, the smaller grains impose (i) a larger frequency and field amplitudes to trigger a heat release, as observed in the experimental results; and (ii) a heat release that balances the benefits of deeper pinning landscapes with the detrimental effect of weaker anisotropy landscape. \newline

We have shown how nanoflowers with larger grains within the flux-closure dominated regime deliver heat comparable to smaller commercial counterparts~\cite{jefremovas2021nanoflowers}, with the key advantage of being less prompt to agglomeration. Samples with smaller grains, despite offering a larger pinning potential in terms of inter-grain boundaries, loose their heating power by offering a smoother anisotropy landscape, highlighting the dominant role of the latter over the inter-grain pinning in the heating response. Furthermore, our experiments allowed to underscore the role of thermal fluctuations, which do not change the qualitative trends, yet they effectively reduce the energy demands on the magnetic field excitation, observing larger heat release at smaller frequency and field amplitudes compared to our micromagnetic athermal simulations.\newline

Our simulations indicate a larger heating performance of vortex core-dominated NFs compared to the flux-closure-dominated counterparts, supporting the $d = 100-120~\mathrm{nm}$ size windows for the hyperthermia ``sweet spot'' \cite{jefremovas2026coercivity}. Both our experiments and simulations reveal a larger energy dissipated for realizations with smaller grains, consequence of the more dense pinning landscape, providing a guide for the rational design of optimized nanotransducers. Our simulations further highlight larger grains yielding a more localized distribution of ``hot spots'' in both space and time, whereas smaller grains spread dissipation over broader regions of the particle and over longer time windows. This spatio-temporal resolution provides a practical design handle: it resolves at intra-particle level where heat is generated and for how long the release is sustained, thereby enabling informed choices of microstructure depending on whether rapid, concentrated heating or more distributed, sustained dissipation is desired.\newline

We can anticipate that, for MHT applications, slower magnetization dynamics imposed by a rich pinning landscape leading to a sustained heat release distributed over time is preferred over short-lived, intense peaks. Such transient power spikes may occur too rapidly to effectively disrupt the metabolic activity of tumoral cells, whereas a more gradual heat release promotes a homogeneous and controllable thermal response, which is particularly desirable for magnetic hyperthermia applications.\newline

\section*{Acknowledgements}
E.M.J. acknowledges funding from the European Union's Horizon 2020 research and innovation program under the Marie Sk{\l}odowska-Curie actions grant agreement No.~101081455-YIA (HYPUSH); the Institute for Advanced Studies (IAS) of the University of Luxembourg; and the Fonds Wetenschappelijk Onderzoek (FWO-Vlaanderen) project number No.~V501325N. P. R. acknowledges Flanders Innovation \& Entrepreneurship (Project No. HBC.2022.0682). The PTB guest scientist program giving the opportunity to P. Rooms to carry
out AC measurements is kindly acknowledged. COST Action CA23132 ``Magnetic Particle Imaging for next-generation theranostics and medical research'' (NexMPI), supported by COST (European Cooperation in Science and Technology) is kindly acknowledged. The computational resources and services used in this work were provided by the HPC facilities of the University of Luxembourg, the Luxembourg national supercomputer MeluXina, the VSC (Flemish Supercomputer Center), funded by the Research Foundation Flanders (FWO) and the Flemish Government – department EWI, and the Donostia International Physics Center (DIPC). The authors gratefully acknowledge the HPC UniLu, LuxProvide, VSC and DIPC Supercomputing Center teams for their technical and human support.

\section*{Author contributions statement}
Conceptualization: E.M.J. Methodology: E.M.J., J. L. Software: E.M.J., J.L. Validation: all authors. Formal analysis: E.M.J., P.R. Investigation: E.M.J., P.R. Sample preparation: A.G-C., MP.M. Resources: All authors. Data curation: E.M.J., P.R. Visualization: E.M.J., P.R. Supervision: A. M., F.W., MP.M., J.L. Project administration: E.M.J., A.M., J.L. Funding acquisition: E.M.J. Writing original draft: E.M.J. Writing, review, and editing: all authors.
\section*{Competing interests}
The authors declare no competing interests.
The corresponding author is responsible for submitting a \href{http://www.nature.com/srep/policies/index.html#competing}{competing interests statement} on behalf of all authors of the paper. This statement must be included in the submitted article file.

\bibliography{main}

@article{rizzo2026gram,
  title={Gram-Scale Production of Iron Oxide Rubik-Cube Nanoparticles: New Tools for the Clinical Translation of Magnetic Hyperthermia and Magnetic Particle Imaging},
  author={Rizzo, Giusy MR and Silvestri, Niccol{\`o} and Jarmouni, Nabila and Gavil{\'a}n, Helena and Yakubu, Hamza and Arenas-Esteban, Daniel and Bals, Sara and Parlanti, Paola and Gemmi, Mauro and Prado, Ross Clark and others},
  journal={Advanced Functional Materials},
  pages={e22732},
  year={2026},
  publisher={Wiley Online Library}
}

@article{allia2011dynamic,
  title={Dynamic effects of dipolar interactions on the magnetic behavior of magnetite nanoparticles},
  author={Allia, Paolo and Tiberto, Paola},
  journal={Journal of Nanoparticle Research},
  volume={13},
  number={12},
  pages={7277--7293},
  year={2011},
  publisher={Springer}
}

@article{blanco2016real,
  title={Real-time tracking of delayed-onset cellular apoptosis induced by intracellular magnetic hyperthermia},
  author={Blanco-Andujar, Cristina and Ortega, Daniel and Southern, Paul and Nesbitt, Stephen A and Thanh, Nguyễn Thị Kim and Pankhurst, Quentin A},
  journal={Nanomedicine},
  volume={11},
  number={2},
  pages={121--136},
  year={2016},
  publisher={Taylor \& Francis}
}

@article{mekseriwattana2025shape,
  title={Shape-Control in Microwave-Assisted Synthesis: A Fast Route to Size-Tunable Iron Oxide Nanocubes with Benchmark Magnetic Heat Losses},
  author={Mekseriwattana, Wid and Silvestri, Niccol{\`o} and Brescia, Rosaria and Tiryaki, Ecem and Barman, Jugal and Mohammadzadeh, Farshad Gorji and Jarmouni, Nabila and Pellegrino, Teresa},
  journal={Advanced Functional Materials},
  volume={35},
  number={3},
  pages={2413514},
  year={2025},
  publisher={Wiley Online Library}
}

@article{hugounenq2012iron,
  title={Iron oxide monocrystalline nanoflowers for highly efficient magnetic hyperthermia},
  author={Hugounenq, Pierre and Levy, Michael and Alloyeau, Damien and Lartigue, Lena{\"\i}c and Dubois, Emmanuelle and Cabuil, Valérie and Ricolleau, Christian and Roux, Stéphane and Wilhelm, Claire and Gazeau, Florence and others},
  journal={The {J}ournal of {P}hysical {C}hemistry {C}},
  volume={116},
  number={29},
  pages={15702--15712},
  year={2012},
  publisher={ACS Publications}
}

@article{collins2007imagej,
  title={ImageJ for microscopy},
  author={Collins, Tony J},
  journal={Biotechniques},
  volume={43},
  number={sup1},
  pages={S25--S30},
  year={2007},
  publisher={Taylor \& Francis}
}

@article{guivar2014structural,
  title={Structural and magnetic properties of monophasic maghemite ($\gamma$-Fe$_{2}$O$_{3}$) nanocrystalline powder},
  author={Guivar, Juan Adri{\'a}n Ramos and Mart{\'\i}nez, Arturo Isa{\'\i}as and Anaya, Ana Osorio and Valladares, Luis De Los Santos and F{\'e}lix, Lizbet Le{\'o}n and Dominguez, Angel Bustamante},
  journal={Advances in Nanoparticles},
  volume={2014},
  year={2014},
  publisher={Scientific Research Publishing}
}

@article{ccam2026tuning,
  title={Tuning field amplitude to minimise heat-loss variability in magnetic hyperthermia},
  author={{\c{C}}am, Necda and L{\'o}pez-V{\'a}zquez, Iago and Iglesias, {\`O}scar and Serantes, David},
  journal={arXiv preprint arXiv:2601.20315},
  year={2026}
}

@article{diaz2025preclinical,
  title={Preclinical Development of Magnetic Nanoparticles for Hyperthermia Treatment of Pancreatic Cancer},
  author={D{\'\i}az-Riascos, Zamira V and Llaguno-Munive, Monserrat and Lafuente-G{\'o}mez, Nuria and Luengo, Yurena and Holmes, Sarah and Volatron, Jeanne and Ibarrola, Oihane and Mancilla, Sandra and Sarno, Francesca and Aguirre, Jos{\'e} Javier and others},
  journal={ACS Applied Materials \& Interfaces},
  year={2025},
  publisher={ACS Publications}
}

@article{prajapati2024transforming,
  title={Transforming cancer detection and treatment with nanoflowers},
  author={Prajapati, Bhupendra G and Verma, Kanika and Sharma, Swapnil and Kapoor, Devesh U},
  journal={Medical Oncology},
  volume={41},
  number={11},
  pages={295},
  year={2024},
  publisher={Springer}
}

@article{wu2024roadmap,
  title={Roadmap on magnetic nanoparticles in nanomedicine},
  author={Wu, Kai and Wang, Jian-Ping and Natekar, Niranjan A and Ciannella, Stefano and Gonz{\'a}lez-Fern{\'a}ndez, Cristina and Gomez-Pastora, Jenifer and Bao, Yuping and Liu, Jinming and Liang, Shuang and Wu, Xian and others},
  journal={Nanotechnology},
  year={2024}
}

@article{riedinger2013subnanometer,
  title={Subnanometer local temperature probing and remotely controlled drug release based on azo-functionalized iron oxide nanoparticles},
  author={Riedinger, Andreas and Guardia, Pablo and Curcio, Alberto and Garcia, Miguel A and Cingolani, Roberto and Manna, Liberato and Pellegrino, Teresa},
  journal={Nano Letters},
  volume={13},
  number={6},
  pages={2399--2406},
  year={2013},
  publisher={ACS Publications}
}

@article{ortega2023estimating,
  title={Estimating the heating of complex nanoparticle aggregates for magnetic hyperthermia},
  author={Ortega-Julia, Javier and Ortega, Daniel and Leliaert, Jonathan},
  journal={Nanoscale},
  volume={15},
  number={24},
  pages={10342--10350},
  year={2023},
  publisher={Royal Society of Chemistry}
}

@article{guibert2015hyperthermia,
  title={Hyperthermia of magnetic nanoparticles: experimental study of the role of aggregation},
  author={Guibert, Cl{\'e}ment and Dupuis, Vincent and Peyre, V{\'e}ronique and Fresnais, J{\'e}r{\^o}me},
  journal={The Journal of Physical Chemistry C},
  volume={119},
  number={50},
  pages={28148--28154},
  year={2015},
  publisher={ACS Publications}
}

@article{khandhar2011monodispersed,
  title={Monodispersed magnetite nanoparticles optimized for magnetic fluid hyperthermia: Implications in biological systems},
  author={Khandhar, Amit P and Ferguson, R Matthew and Krishnan, Kannan M},
  journal={Journal of applied physics},
  volume={109},
  number={7},
  year={2011},
  publisher={AIP Publishing}
}

@article{jeun2009effects,
  title={Effects of particle dipole interaction on the ac magnetically induced heating characteristics of ferrite nanoparticles for hyperthermia},
  author={Jeun, Minhong and Bae, Seongtae and Tomitaka, Asahi and Takemura, Yasushi and Park, Ki Ho and Paek, Sun Ha and Chung, Kyung-Won},
  journal={Applied Physics Letters},
  volume={95},
  number={8},
  year={2009},
  publisher={AIP Publishing}
}

@article{beola2020intracellular,
  title={The intracellular number of magnetic nanoparticles modulates the apoptotic death pathway after magnetic hyperthermia treatment},
  author={Beola, Lilianne and As{\'\i}n, Laura and Roma-Rodrigues, Catarina and Fern{\'a}ndez-Afonso, Yilian and Fratila, Raluca M and Serantes, David and Ruta, Sergiu and Chantrell, Roy W and Fernandes, Alexandra R and Baptista, Pedro V and others},
  journal={ACS Applied Materials \& Interfaces},
  volume={12},
  number={39},
  pages={43474--43487},
  year={2020},
  publisher={ACS Publications}
}

@article{hergt2009validity,
  title={Validity limits of the N{\'e}el relaxation model of magnetic nanoparticles for hyperthermia},
  author={Hergt, Rudolf and Dutz, Silvio and Zeisberger, Matthias},
  journal={Nanotechnology},
  volume={21},
  number={1},
  pages={015706},
  year={2009},
  publisher={IOP Publishing}
}

@article{hergt2007magnetic,
  title={Magnetic particle hyperthermia—biophysical limitations of a visionary tumour therapy},
  author={Hergt, Rudolf and Dutz, Silvio},
  journal={Journal of Magnetism and Magnetic Materials},
  volume={311},
  number={1},
  pages={187--192},
  year={2007},
  publisher={Elsevier}
}

@article{gavilan2021magnetic,
  title={Magnetic nanoparticles and clusters for magnetic hyperthermia: Optimizing their heat performance and developing combinatorial therapies to tackle cancer},
  author={Gavil{\'a}n, Helena and Avugadda, Sahitya Kumar and Fern{\'a}ndez-Cabada, Tamara and Soni, Nisarg and Cassani, Marco and Mai, Binh T and Chantrell, Roy and Pellegrino, Teresa},
  journal={Chemical Society Reviews},
  volume={50},
  number={20},
  pages={11614--11667},
  year={2021},
  publisher={Royal Society of Chemistry}
}

@article{vivas2020toward,
  title={Toward understanding complex spin textures in nanoparticles by magnetic neutron scattering},
  author={Vivas, Laura G and Yanes, Rocio and Berkov, Dmitry and Erokhin, Sergey and Bersweiler, Mathias and Honecker, Dirk and Bender, Philipp and Michels, Andreas},
  journal={Physical Review Letters},
  volume={125},
  number={11},
  pages={117201},
  year={2020},
  publisher={APS}
}

@article{lappas2019vacancy,
  title={Vacancy-Driven Noncubic Local Structure and Magnetic Anisotropy Tailoring in Fe\textsubscript{x}O\textsuperscript{-}Fe\textsubscript{3-$\delta$}O\textsubscript{4} Nanocrystals},
  author={Lappas, Alexandros and Antonaropoulos, George and Brintakis, Konstantinos and Vasilakaki, Marianna and Trohidou, Kalliopi N and Iannotti, Vincenzo and Ausanio, Giovanni and Kostopoulou, Athanasia and Abeykoon, Milinda and Robinson, Ian K and others},
  journal={Physical Review X},
  volume={9},
  number={4},
  pages={041044},
  year={2019},
  publisher={APS}
}

@book{coey2010magnetism,
  title={Magnetism and magnetic materials},
  author={Coey, John MD},
  year={2010},
  publisher={Cambridge {U}niversity {P}ress}
}

@article{serantes2021nanoparticle,
  title={Nanoparticle size threshold for magnetic agglomeration and associated hyperthermia performance},
  author={Serantes, David and Baldomir, Daniel},
  journal={Nanomaterials},
  volume={11},
  number={11},
  pages={2786},
  year={2021},
  publisher={MDPI}
}

@article{Vansteenkiste2014,
    author  = {Vansteenkiste, Arne and
               Leliaert, Jonathan and
               Dvornik, Mykola and
               Helsen, Mathias and
               Garcia-Sanchez, Felipe and
               {Van Waeyenberge}, Bartel},
    title   = {{The design and verification of Mumax3}},
    journal = {AIP Advances},
    number  = {10},
    pages   = {107133},
    volume  = {4},
    year    = {2014},
    doi     = {10.1063/1.4899186},
    url     = {http://doi.org/10.1063/1.4899186}
}

@article{etheridge2014accounting,
  title={Accounting for biological aggregation in heating and imaging of magnetic nanoparticles},
  author={Etheridge, Michael L and Hurley, Katie R and Zhang, Jinjin and Jeon, Seongho and Ring, Hattie L and Hogan, Christopher and Haynes, Christy L and Garwood, Michael and Bischof, John C},
  journal={Technology},
  volume={2},
  number={03},
  pages={214--228},
  year={2014},
  publisher={World Scientific}
}

@article{Lel2014,
    author  = {Leliaert, Jonathan and
               Van de Wiele, Ben and
               Vansteenkiste, Arne and
               Laurson, Lasse and
               Durin, Gianfranco and
               Dupr{\'e}, Luc and
               Van Waeyenberge, Bartel},
    title   = {{Current-driven domain wall mobility in polycrystalline permalloy nanowires: A numerical study}},
    journal = {Journal of Applied Physics},
    volume  = {115},
    number  = {23},
    pages   = {233903},
    year    = {2014},
    doi     = {10.1063/1.4883297},
    url     = {http://dx.doi.org/10.1063/1.4883297}
}

@article{moya2024unveiling,
  title={Unveiling the crystal and magnetic texture of iron oxide nanoflowers},
  author={Moya, Carlos and Escoda-Torroella, Mariona and Rodr{\'\i}guez-{\'A}lvarez, Javier and Figueroa, Adriana I and Garc{\'\i}a, {\'I}ker and Ferrer-Vidal, In{\'e}s Batalla and Gallo-Cordova, A and Morales, M Puerto and Aballe, Luc{\'\i}a and Rodr{\'\i}guez, Arantxa Fraile and others},
  journal={Nanoscale},
  volume={16},
  number={4},
  pages={1942--1951},
  year={2024},
  publisher={Royal Society of Chemistry}
}

@article{di2012generalization,
  title={A generalization of the fundamental theorem of Brown for fine ferromagnetic particles},
  author={Di Fratta, Giovanni and Serpico, Claudio and d'Aquino, Massimiliano},
  journal={Physica B},
  volume={407},
  number={9},
  pages={1368--1371},
  year={2012},
  publisher={Elsevier}
}

@article{pankhurst2003applications,
  title={Applications of magnetic nanoparticles in biomedicine},
  author={Pankhurst, Quentin A and Connolly, Jon and Jones, Stephen K and Dobson, JJJ},
  journal={Journal of physics D: Applied Physics},
  volume={36},
  number={13},
  pages={R167},
  year={2003},
  publisher={IOP Publishing}
}

@article{gavilan2017formation,
  title={Formation mechanism of maghemite nanoflowers synthesized by a polyol-mediated process},
  author={Gavil{\'a}n, Helena and S{\'a}nchez, Elena H. and Brollo, Mar{\'i}a E. F. and As{\'i}n, Laura and Moerner, Kimmie K. and Frandsen, Cathrine and L{\'a}zaro, Francisco J. and Serna, Carlos J. and Veintemillas-Verdaguer, Sabino and Morales, Mar{\'i}a Puerto and Guti{\'e}rrez, Luc{\'i}a},
  journal={ACS Omega},
  volume={2},
  number={10},
  pages={7172--7184},
  year={2017},
  publisher={ACS Publications}
}

@book{CullityStock2001,
  title     = {Elements of X-Ray Diffraction},
  author    = {Cullity, B. D. and Stock, S. R.},
  edition   = {3},
  publisher = {Prentice Hall / Pearson},
  year      = {2001},
  address   = {Upper Saddle River, NJ},
  isbn      = {978-0201610918}
}

@article{gallo2024magnetic,
  title={Magnetic harvesting and degradation of microplastics using iron oxide nanoflowers prepared by a scaled-up procedure},
  author={Gallo-Cordova, Alvaro and Corrales-P{\'e}rez, Bel{\'e}n and Cabrero, Paula and Force, Carmen and Veintemillas-Verdaguer, Sabino and Ovejero, Jes{\'u}s G and P.  Morales, Mar{\'i}a},
  journal={Chemical Engineering Journal},
  volume={490},
  pages={151725},
  year={2024},
  publisher={Elsevier}
}

@article{kumar2011magnetic,
  title={Magnetic nanomaterials for hyperthermia-based therapy and controlled drug delivery},
  author={Kumar, Challa SSR and Mohammad, Faruq},
  journal={Advanced drug delivery reviews},
  volume={63},
  number={9},
  pages={789--808},
  year={2011},
  publisher={Elsevier}
}

@article{perigo2015fundamentals,
  title={Fundamentals and advances in magnetic hyperthermia},
  author={Perigo, Elio Alberto and Hemery, Gauvin and Sandre, Olivier and Ortega, Daniel and Garaio, Eneko and Plazaola, Fernando and Teran, Francisco Jose},
  journal={Applied Physics Reviews},
  volume={2},
  number={4},
  year={2015},
  publisher={AIP Publishing}
}

@article{gavilan2025magnetic,
  title={Magnetic hyperthermia in focus: emerging non-cancer applications of magnetic nanoparticles},
  author={Gavil{\'a}n, Helena and Gallo-Cordova, Alvaro and Chediak, Maura Lisett R{\'a}bade and Rod{\'\i}guez, Amira P{\'a}ez and  Morales, Maria P. and Guti{\'e}rrez, Luc{\'i}a},
  journal={Nanoscale},
  volumen={17},
  pages={27734-27761},
  year={2025},
  publisher={Royal Society of Chemistry}
}

@article{rivera2021emerging,
  title={Emerging biomedical applications based on the response of magnetic nanoparticles to time-varying magnetic fields},
  author={Rivera-Rodriguez, Angelie and Rinaldi-Ramos, Carlos M},
  journal={Annual Review of Chemical and Biomolecular Engineering},
  volume={12},
  pages={163--185},
  year={2021},
  publisher={Annual Reviews}
}

@article{colombo2012biological,
  title={Biological applications of magnetic nanoparticles},
  author={Colombo, Miriam and Carregal-Romero, Susana and Casula, Maria F and Guti{\'e}rrez, Luc{\'i}a and Morales, Mar{\'i}a P and B{\"o}hm, Ingrid B and Heverhagen, Johannes T and Prosperi, Davide and Parak, Wolfgang J},
  journal={Chemical Society Reviews},
  volume={41},
  number={11},
  pages={4306--4334},
  year={2012},
  publisher={Royal Society of Chemistry}
}

@article{pankhurst2009progress,
  title={Progress in applications of magnetic nanoparticles in biomedicine},
  author={Pankhurst, QA and Thanh, NTK and Jones, SK and Dobson, J},
  journal={Journal of Physics D: Applied Physics},
  volume={42},
  number={22},
  pages={224001},
  year={2009},
  publisher={IOP Publishing}
}

@article{goiriena2020disk,
  title={Disk-shaped magnetic particles for cancer therapy},
  author={Goiriena-Goikoetxea, M and Mu{\~n}oz, D and Orue, I and Fern{\'a}ndez-Gubieda, ML and Bokor, J and Muela, A and Garc{\'\i}a-Arribas, A},
  journal={Applied Physics Reviews},
  volume={7},
  number={1},
  year={2020},
  publisher={AIP Publishing}
}

@article{goiriena2016high,
  title={High-yield fabrication of 60 nm Permalloy nanodiscs in well-defined magnetic vortex state for biomedical applications},
  author={Goiriena-Goikoetxea, Maite and Garc{\'\i}a-Arribas, A and Rouco, M and Svalov, AV and Barandiaran, JM},
  journal={Nanotechnology},
  volume={27},
  number={17},
  pages={175302},
  year={2016},
  publisher={IOP Publishing}
}

@article{ho2011monodisperse,
  title={Monodisperse magnetic nanoparticles for theranostic applications},
  author={Ho, DON and Sun, Xiaolian and Sun, Shouheng},
  journal={Accounts of Chemical Research},
  volume={44},
  number={10},
  pages={875--882},
  year={2011},
  publisher={ACS Publications}
}

@article{simeonidis2020controlling,
  title={Controlling magnetization reversal and hyperthermia efficiency in core--shell iron--iron oxide magnetic nanoparticles by tuning the interphase coupling},
  author={Simeonidis, Konstantinos and Martinez-Boubeta, Carlos and Serantes, David and Ruta, S and Chubykalo-Fesenko, O and Chantrell, R and Or{\'o}-Sol{\'e}, J and Balcells, Ll and Kamzin, AS and Nazipov, RA and others},
  journal={ACS applied nano materials},
  volume={3},
  number={5},
  pages={4465--4476},
  year={2020},
  publisher={ACS Publications}
}

@article{myrovali2023toward,
  title={Toward the separation of different heating mechanisms in magnetic particle hyperthermia},
  author={Myrovali, Eirini and Papadopoulos, Kyrillos and Charalampous, Georgia and Kesapidou, Paraskevi and Vourlias, George and Kehagias, Thomas and Angelakeris, Makis and Wiedwald, Ulf},
  journal={ACS Omega},
  volume={8},
  number={14},
  pages={12955--12967},
  year={2023},
  publisher={ACS Publications}
}

@article{liu2020comprehensive,
  title={Comprehensive understanding of magnetic hyperthermia for improving antitumor therapeutic efficacy},
  author={Liu, Xiaoli and Zhang, Yifan and Wang, Yanyun and Zhu, Wenjing and Li, Galong and Ma, Xiaowei and Zhang, Yihan and Chen, Shizhu and Tiwari, Shivani and Shi, Kejian and others},
  journal={Theranostics},
  volume={10},
  number={8},
  pages={3793},
  year={2020}
}

@article{coisson2017hysteresis,
  title={Hysteresis losses and specific absorption rate measurements in magnetic nanoparticles for hyperthermia applications},
  author={Co{\"\i}sson, Marco and Barrera, Gabriele and Celegato, Federica and Martino, Luca and Kane, Shashank N and Raghuvanshi, Saroj and Vinai, Franco and Tiberto, Paola},
  journal={Biochimica et Biophysica Acta (BBA)-General Subjects},
  volume={1861},
  number={6},
  pages={1545--1558},
  year={2017},
  publisher={Elsevier}
}

@article{borchers2025magnetic,
  title={Magnetic Anisotropy Dominates over Physical and Magnetic Structure in Performance of Magnetic Nanoflowers},
  author={Borchers, Julie and Krycka, Kathryn and Bosch-Santos, Brianna and de Lima Correa, Eduardo and Sharma, Anirudh and Carlton, Hayden and Dang, Yanliu and Donahue, Michael and Gr{\"u}ttner, Cordula and Ivkov, Robert and others},
  journal={Small Structures},
  volume={6},
  number={2},
  pages={2400410},
  year={2025},
  publisher={Wiley Online Library}
}

@article{jefremovas2021nanoflowers,
  title={Nanoflowers versus magnetosomes: comparison between two promising candidates for magnetic hyperthermia therapy},
  author={Jefremovas, Elizabeth M and Gandarias, Luc{\'\i}a and Rodrigo, Irati and Marcano, Lourdes and Gr{\"u}ttner, Cordula and Garc{\'\i}a, Jos{\'e} {\'A}ngel and Garayo, Eneko and Orue, I{\~n}aki and Garc{\'\i}a-Prieto, Ana and Muela, Alicia and others},
  journal={IEEE Access},
  volume={9},
  pages={99552--99561},
  year={2021},
  publisher={IEEE}
}

@article{bender2018relating,
  title={Relating magnetic properties and high hyperthermia performance of iron oxide nanoflowers},
  author={Bender, Philipp and Fock, Jeppe and Frandsen, Cathrine and Hansen, Mikkel F and Balceris, Christoph and Ludwig, Frank and Posth, Oliver and Wetterskog, Erik and Bogart, Lara K and Southern, Paul and others},
  journal={The Journal of Physical Chemistry C},
  volume={122},
  number={5},
  pages={3068--3077},
  year={2018},
  publisher={ACS Publications}
}

@article{shokrollahi2017review,
  title={A review of the magnetic properties, synthesis methods and applications of maghemite},
  author={Shokrollahi, HJJOM},
  journal={Journal of Magnetism and Magnetic Materials},
  volume={426},
  pages={74--81},
  year={2017},
  publisher={Elsevier}
}

@article{roca2007effect,
  title={Effect of nature and particle size on properties of uniform magnetite and maghemite nanoparticles},
  author={Roca, Alejandro G and Marco, Jose F and Morales, Mar{\'\i}a P.  and Serna, Carlos J},
  journal={The Journal of Physical Chemistry C},
  volume={111},
  number={50},
  pages={18577--18584},
  year={2007},
  publisher={ACS Publications}
}

@article{gavilan2021size,
  title={How size, shape and assembly of magnetic nanoparticles give rise to different hyperthermia scenarios},
  author={Gavil{\'a}n, Helena and Simeonidis, Konstantinos and Myrovali, Eirini and Mazar{\'\i}o, Eva and Chubykalo-Fesenko, Oksana and Chantrell, R and Balcells, Ll and Angelakeris, Mavroidis and Morales, M Puerto and Serantes, David},
  journal={Nanoscale},
  volume={13},
  number={37},
  pages={15631--15646},
  year={2021},
  publisher={Royal Society of Chemistry}
}

@article{gross2021magnetic,
  title={Magnetic anisotropy of individual maghemite mesocrystals},
  author={Gross, B and Philipp, S and Josten, E and Leliaert, Jonathan and Wetterskog, Erik and Bergstr{\"o}m, Lennart and Poggio, M},
  journal={Physical Review B},
  volume={103},
  number={1},
  pages={014402},
  year={2021},
  publisher={APS}
}

@article{pisane2017unusual,
  title={Unusual enhancement of effective magnetic anisotropy with decreasing particle size in maghemite nanoparticles},
  author={Pisane, KL and Singh, Sobhit and Seehra, MS},
  journal={Applied Physics Letters},
  volume={110},
  number={22},
  year={2017},
  publisher={AIP Publishing}
}

@article{sinaga2024neutron,
  title={Neutron scattering signature of the Dzyaloshinskii-Moriya interaction in nanoparticles},
  author={Sinaga, Evelyn Pratami and Adams, Michael P and Hasdeo, Eddwi H and Michels, Andreas},
  journal={Physical Review B},
  volume={110},
  number={5},
  pages={054404},
  year={2024},
  publisher={APS}
}

@article{gavilán2017formation,
  title={Formation mechanism of maghemite nanoflowers synthesized by a polyol-mediated process},
  author={Gavil\'an, Helena and S\'anchez, Elena H and Brollo, Mar{\'\i}a EF and As{\'\i}n, Laura and Moerner, Kimmie K and Frandsen, Cathrine and L\'azaro, Francisco J and Serna, Carlos J and Veintemillas-Verdaguer, Sabino and Morales, M Puerto and others},
  journal={ACS Omega},
  volume={2},
  number={10},
  pages={7172--7184},
  year={2017},
  publisher={ACS Publications}
}

@article{usov2018magnetic,
  title={Magnetic vortices as efficient nano heaters in magnetic nanoparticle hyperthermia},
  author={Usov, NA and Nesmeyanov, MS and Tarasov, VP},
  journal={Scientific Reports},
  volume={8},
  number={1},
  pages={1224},
  year={2018},
  publisher={Nature Publishing Group UK London}
}

@article{jefremovas2026coercivity,
  title={Coercivity-Size Map of Magnetic Nanoflowers: Spin Disorder Tunes the Vortex Reversal Mechanism and Tailors the Hyperthermia Sweet Spot},
  author={Jefremovas, Elizabeth M and Calus, Lisa and Leliaert, Jonathan},
  journal={Small Science},
  volume={6},
  number={1},
  pages={e202500490},
  year={2026},
  publisher={Wiley Online Library}
}

@article{munoz2020disentangling,
  title={Disentangling local heat contributions in interacting magnetic nanoparticles},
  author={Mu{\~n}oz-Menendez, Cristina and Serantes, David and Chubykalo-Fesenko, Oksana and Ruta, Sergiu and Hovorka, Ondrej and Nieves, Pablo and Livesey, KL and Baldomir, Daniel and Chantrell, R},
  journal={Physical Review B},
  volume={102},
  number={21},
  pages={214412},
  year={2020},
  publisher={APS}
}

@article{leliaert2021individual,
  title={Individual particle heating of interacting magnetic nanoparticles at nonzero temperature},
  author={Leliaert, Jonathan and Ortega-Julia, Javier and Ortega, Daniel},
  journal={Nanoscale},
  volume={13},
  number={35},
  pages={14734--14744},
  year={2021},
  publisher={Royal Society of Chemistry}
}

\end{document}


\title{Supplemental Material to: Nanoscale mapping of internal magnetization dynamics reveals how disorder shapes heat generation in magnetic particle hyperthermia }
\author{Elizabeth M Jefremovas}
\email{elizabeth.jefremovas@.uni.lu}
\affiliation{Department of Physics and Materials Science, University of Luxembourg, 162A Avenue de la Faiencerie, L-1511 Luxembourg, Grand Duchy of Luxembourg}
\affiliation{Institute for Advanced Studies, University of Luxembourg, Campus Belval, L-4365 Esch-sur-Alzette, Luxembourg}
\author{Pauline Rooms}
\affiliation{DyNaMat, Department of Solid State Sciences, Ghent University, 9000 Ghent, Belgium}
\author{\'Alvaro Gallo-C\'ordova}
\affiliation{Instituto de Ciencia de Materiales de Madrid (ICMM-CSIC), Madrid 28049, Spain}
\author{Mar\'ia P. Morales}
\affiliation{Instituto de Ciencia de Materiales de Madrid (ICMM-CSIC), Madrid 28049, Spain}
\author{Frank Wiekhorst}
\affiliation{Physikalisch-Technische Bundesanstalt, D-10587 Berlin, Germany}
\author{Andreas Michels}
\affiliation{Department of Physics and Materials Science, University of Luxembourg, 162A Avenue de la Faiencerie, L-1511 Luxembourg, Grand Duchy of Luxembourg}
\author{Jonathan Leliaert}
\affiliation{DyNaMat, Department of Solid State Sciences, Ghent University, 9000 Ghent, Belgium}

\date{\today}

\begin{abstract}
This Supplemental Material includes the structural characterization of the experimentally measured nanoflowers (transmission electron microscopy and X-ray diffraction); extended discussions on the validity of our simulations without thermal noise; experimental AC hysteresis loops, from which the $SAR$ is extracted; a linear extrapolation of the experimentally-measured $SAR$ at $f = 100~\mathrm{kHz}$; temperature mappings and heat release for $(d, g_{s}) = (220, 10)~\mathrm{and}~(250, 25)~\mathrm{nm}$; and temperature mappings, heat release and magnetization reversal of a representative single-domain spherical nanoparticle.
\end{abstract} 

\maketitle
\section{Structural characterization}

\begin{figure*}
\centering
\resizebox{1.0\columnwidth}{!}{\includegraphics{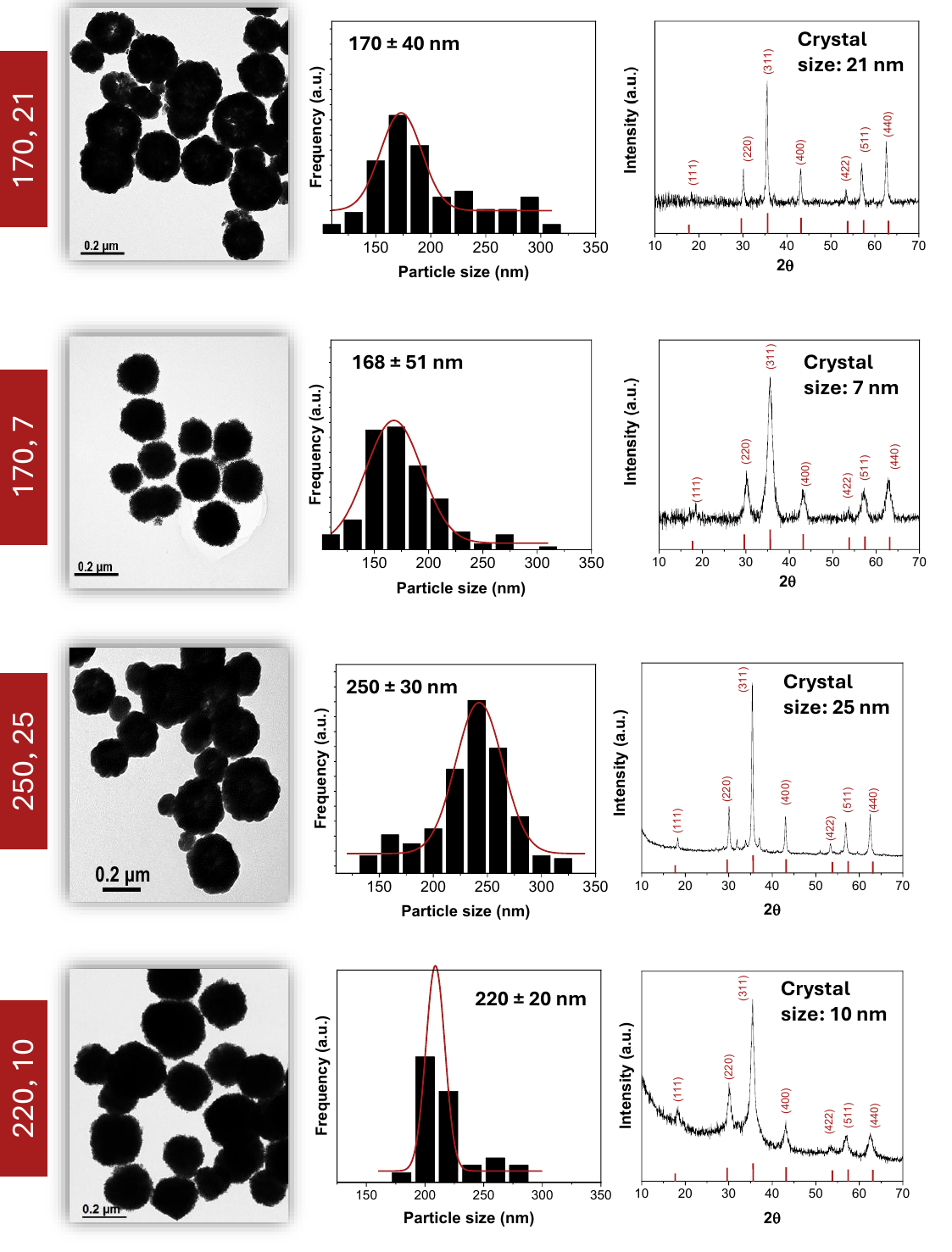}}
    \caption{Representative TEM images, for the four different batches of NFs, along with their size distribution analysis, fitted to a log-normal distribution, and XRD patterns.}
    \label{TEM_XRD}
\end{figure*}

Transmission electron microscopy (TEM) was performed using a JEOL JEM 1010 electron microscope at an accelerating voltage of $100 \, \mathrm{kV}$. For sample preparation, a diluted suspension of NFs was deposited onto an amorphous 300~micron meshed carbon-coated copper grid, followed by evaporation at room temperature. The particle-size distribution was analyzed using the standard software ImageJ~\cite{collins2007imagej} for digital electron microscope image processing, assimilating the particle diameter as the largest internal dimension. Data for an approximate number of at least 200 particles were acquired and fitted to a lognormal distribution, allowing for the calculation of the mean particle size and characterization of the particle size distribution. Figure~\ref{TEM_XRD} includes representative TEM images together with the corresponding statistical results for mean particle size, included also in Table S~\ref{tabla_tamaños}. \newline

To identify the phase and crystalline structure of the NFs, X-ray diffraction (XRD) measurements were performed using a Bruker D8 Advance diffractometer equipped with a graphite monochromator and a Cu-K$\alpha$ source ($\lambda = 1.5406$~\(\text{\AA}\)). The XRD patterns were recorded at room temperature and ambient pressure in the 2$\theta$ range from 20$^{\circ}$ to 70$^{\circ}$. The results, that can be inspected in Figure~\ref{TEM_XRD}, where the most intense peaks have been indexed according to their Miller indices, confirm a single spinel iron-oxide phase, consistent with $\gamma$-Fe$_2$O$_3$/Fe$_3$O$_4$. Considering the post-synthetic oxidative treatment, the phase is assigned to maghemite. To estimate the mean crystal size, corresponding to the size of the grains, Scherrer's equation was used \cite{CullityStock2001},
\begin{equation}
    D = \frac{K\,\lambda}{\beta \cos\theta}
\end{equation}
where $K$ is typically taken as 0.9 for nanoparticles and $\beta$ represents the full width at half-maximum of the most intense peak, which in our case corresponds to the (3 1 1) reflection. The results for grain size are also included in Table S~\ref{tabla_tamaños}. \newline

\begin{table}[h]\label{tabla_tamaños}
\centering
\caption{Structural parameters of the studied samples: Mean diameter $\langle d \rangle$ and corresponding standard deviation $\sigma$, and grain size $g_{s}$. }
\label{tab:samples}
\begin{tabular}{c c c c}
\hline
Sample & $\langle d \rangle$ (nm) & $\sigma$ (nm) & $g_{s}$ (nm) \\
\hline
(170, 21) & 170 & 40 & 21 \\
(170, 7)  & 168 & 51 & 7  \\
(250, 25) & 250 & 20 & 25 \\
(220, 10) & 220 & 30 & 10 \\
\hline
\end{tabular}
\end{table}

\section{Magnetic Fluid Hyperthermia experimental measurements}
\begin{figure*}
    \centering
    \resizebox{1.0\columnwidth}{!}{\includegraphics{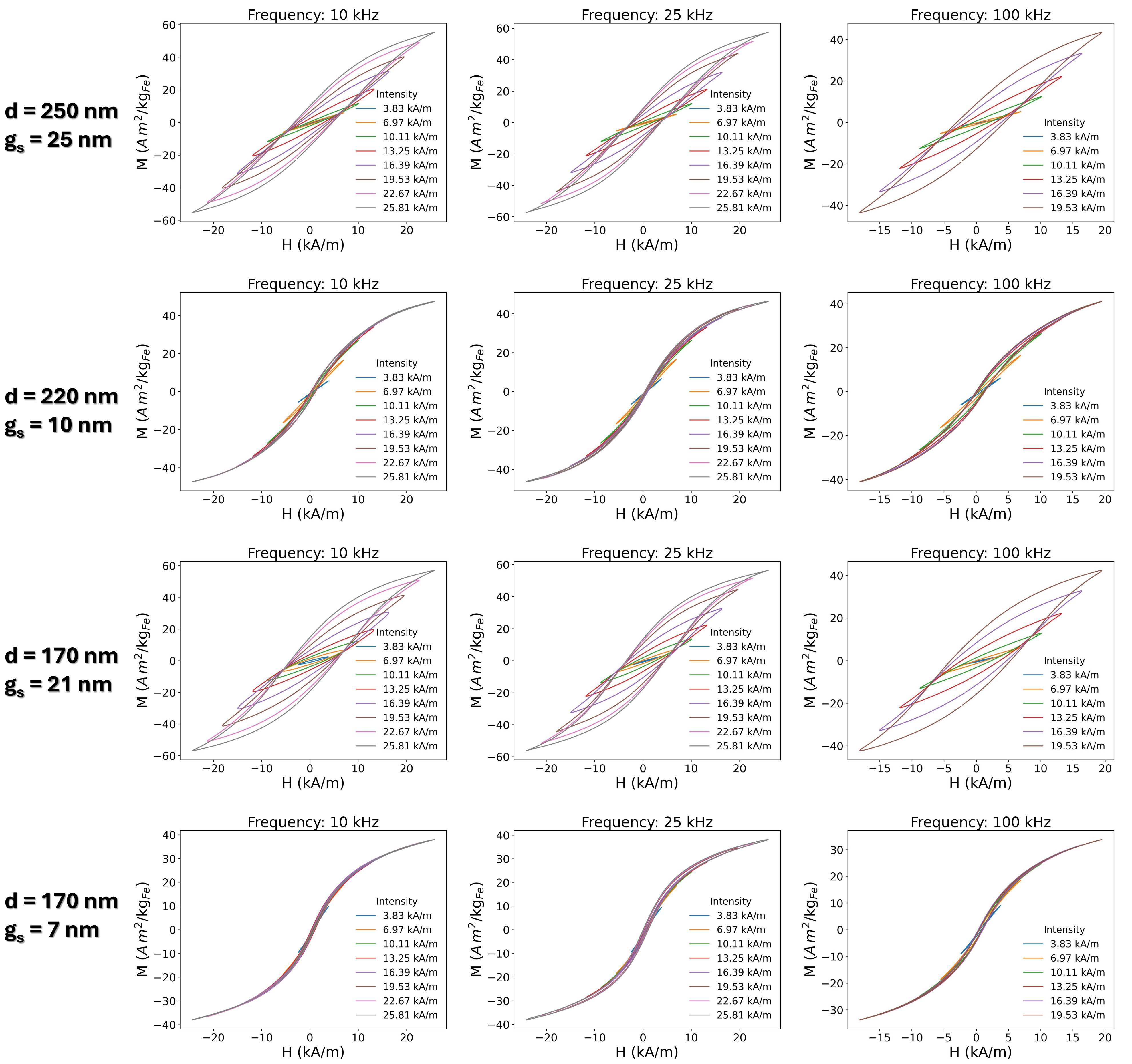}}
    \caption{AC hysteresis loops measured for the NFs with $d = 250, 220~\mathrm{and}~170~\mathrm{nm}$ with grain sizes of $g_{s} = 25, 10, 21~\mathrm{and}~7~\mathrm{nm}$ (top to bottom), at $f$ = 10, 25 and 100 kHz, with AC field amplitudes up to $h_{\text{AC}}$ = 25.81 kA/m, respectively.}
    \label{AC_loops}
\end{figure*}

In order to determine experimentally the heating efficiency of the nanoflowers, we have performed AC hysteresis loops measurements and calculated their Specific Absorption Rate ($SAR$) from the hysteresis losses associated. The AC hysteresis loops have been measured for 4 different NF samples of several diameter and grain sizes: $d = 170~\mathrm{nm}$, with grain sizes of $g_{s} = 21~\mathrm{and}~7~\mathrm{nm}$, and $d = 220~\mathrm{and}~250~\mathrm{nm}$, with $g_{s} = 10~\mathrm{and}~25~\mathrm{nm}$, respectively. The AC loops have been measured at three different frequencies, $f$ = 10, 25, and 100 kHz, with AC field amplitudes from $h_{AC}$ = 3.83 kA/m to 25.81 kA/m~\textit{i.e.}, all within the clinical safety limits $h\cdot f \leq 5~\mathrm{GA/ms}$~\cite{hergt2007magnetic, hergt2009validity}. These loops are included in Fig.~\ref{AC_loops}, where it can be seen how the AC loops become wider at increasing the field amplitude $h_{AC}$, evolving from the narrow and elongated lancet shape at low fields, with low hysteresis losses and associated heating efficiencies, to wider and square-like loops of increasing heat losses.\newline

A more insightful comparison emerges when contrasting samples with smaller grain sizes ($g_{s} = 7$ and $10~\mathrm{nm}$) to those with larger grains ($g_{s} = 21$ and $25~\mathrm{nm}$). While the former exhibit no measurable coercivity under any tested field amplitude or frequency, the latter progressively develop broader and more square-shaped hysteresis loops, reaching $\mu_{0}H_{c} \cong 6.3~\mathrm{mT}$ at applied fields of 16.39~kA/m for $f = 10.0~\mathrm{kHz}$ and 19.53~kA/m for $f = 100.0~\mathrm{kHz}$. This behaviour is illustrated in Fig.~\ref{AC_loops}, where we show the loops acquired at $h_{\mathrm{AC}} = 19.53~\mathrm{kA/m}$ for all four samples across several frequencies.

As seen in the figure, the samples with $g_{s} = 21$ and $25~\mathrm{nm}$ display nearly identical hysteresis loop shapes: wide, square loops with $\mu_{0}H_{c} \cong 6.3~\mathrm{mT}$ and maximum magnetization values of $M \cong 40~\mathrm{A\,m^{2}\,kg^{-1}}$, exhibiting only minor frequency dependence. In contrast, samples with $g_{s} = 7$ and $10~\mathrm{nm}$ generate loops with almost negligible coercivity. Their maximum magnetization also differs: for particles of $d = 220~\mathrm{nm}$, the saturation magnetization is approximately 1.2 times higher than in the $d = 170~\mathrm{nm}$ nanoflowers.

At the highest measured frequency, $f = 100~\mathrm{kHz}$, the small-grain samples display only a marginal coercivity of $\mu_{0}H_{c} \cong 1~\mathrm{mT}$ and $0.2~\mathrm{mT}$, respectively. This indicates that such small grains do not efficiently activate hysteresis losses at sub-hundred-kilohertz frequencies.\newline

\section{Extrapolation of $SAR$ at higher $h_{AC}$ amplitudes}
\begin{figure}
    \centering
    \resizebox{1.0\columnwidth}{!}{\includegraphics{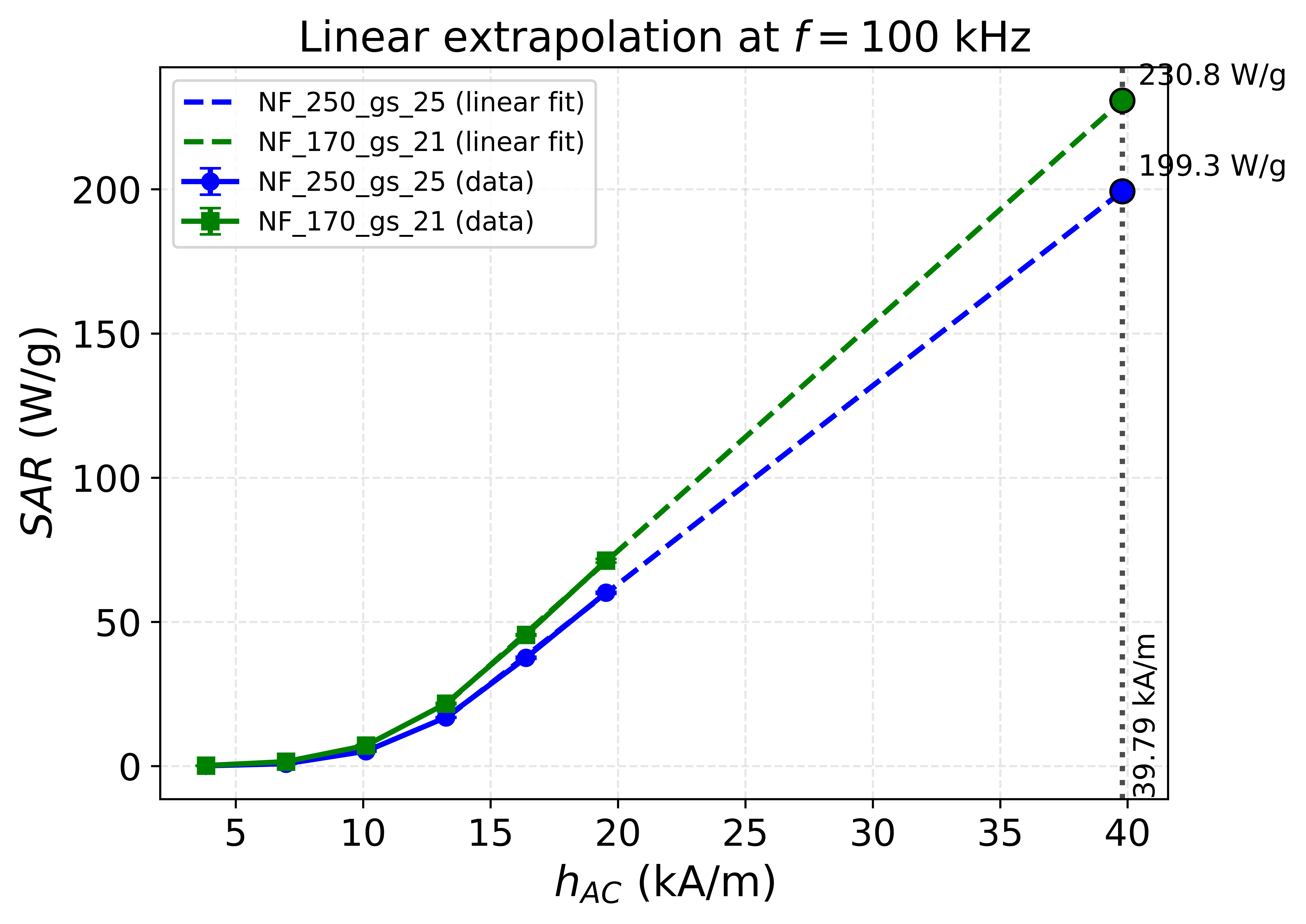}}
    \caption{Specific absorption rate (SAR) as a function of the AC field amplitude $h_{\mathrm{AC}}$ at $f = 100~\mathrm{kHz}$ for $(d, gs) = (250, 25)~\mathrm{nm}~\mathrm{and}~(170, 21)$ NFs. Symbols denote experimental data, while dashed lines correspond to linear fits performed in the linear regime ($h_{\mathrm{AC}} \geq 13.25~\mathrm{kA/m}$). Filled markers indicate extrapolated SAR values at $h_{\mathrm{AC}} = 39.79~\mathrm{kA/m}$ ($\approx 50~\mathrm{mT}$).}
    \label{extrapolation}
\end{figure}

Figure~\ref{extrapolation} includes the experimental data, together with linear fits performed in the linear regime ($h_{\mathrm{AC}} \geq 13.25~\mathrm{kA/m}$), for $(d, gs) = (250, 25)~\mathrm{and}~(170, 21)$ NFs. A $SAR$ value of $\approx 231~\mathrm{W/g}$ and $\approx 199~\mathrm{W/g}$ is estimated, similar to the one obtained for an ensemble of single-domain NFs ($d = 45~\mathrm{nm}$), as reported in Ref.~\cite{jefremovas2021nanoflowers}, yet with a concentration 1.5 smaller for the present study.\newline

\section{Validity of simulations without thermal noise}

Our micromagnetic simulations under frequency-dependent AC magnetic fields were performed without explicitly including a stochastic thermal field associated with finite-temperature fluctuations. This choice is motivated by computational feasibility and by the physical regime addressed in this work. Incorporating a thermal noise term within stochastic Landau--Lifshitz--Gilbert dynamics would dramatically increase the computational cost of the large-scale parametric study presented here, while providing limited additional insight into the dominant magnetization-reversal mechanisms and energy dissipation processes of interest.  \newline

The nanoflowers considered in this study are well above both the superparamagnetic and single-domain limits. The superparamagnetic diameter can be estimated as
\[
d_{\mathrm{SPM}} \approx \left( \frac{6 \cdot 25 \cdot k_B T}{\pi K_u} \right)^{1/3} \approx 27~\mathrm{nm},
\]
assuming $\tau = \tau_0 \exp\!\left( K_u V / k_B T \right)$ with $\ln(\tau/\tau_0)=25$ at room temperature \cite{coey2010magnetism}. The single-domain limit can be approximated as $d_{\mathrm{SD}} \approx 7.2\,l_{\mathrm{ex}}$ \cite{di2012generalization}, corresponding to $\sim 51~\mathrm{nm}$ for maghemite nanoflowers \cite{jefremovas2026coercivity}. Within the investigated size range and excitation frequencies (10--300~kHz), the magnetization dynamics is therefore neither governed by N\'eel nor by Brownian relaxation. Instead, the response evolves from thermally assisted diffusion at low fields to field-induced crossing of energy barriers at larger amplitudes, resulting in a hysteretic regime \cite{coisson2017hysteresis,simeonidis2020controlling,myrovali2023toward,liu2020comprehensive}, where heat release is primarily driven by the applied field with minor influence of thermal fluctuations on the absolute figures. \newline

\begin{figure*}
    \centering
    \resizebox{1.0\columnwidth}{!}{\includegraphics{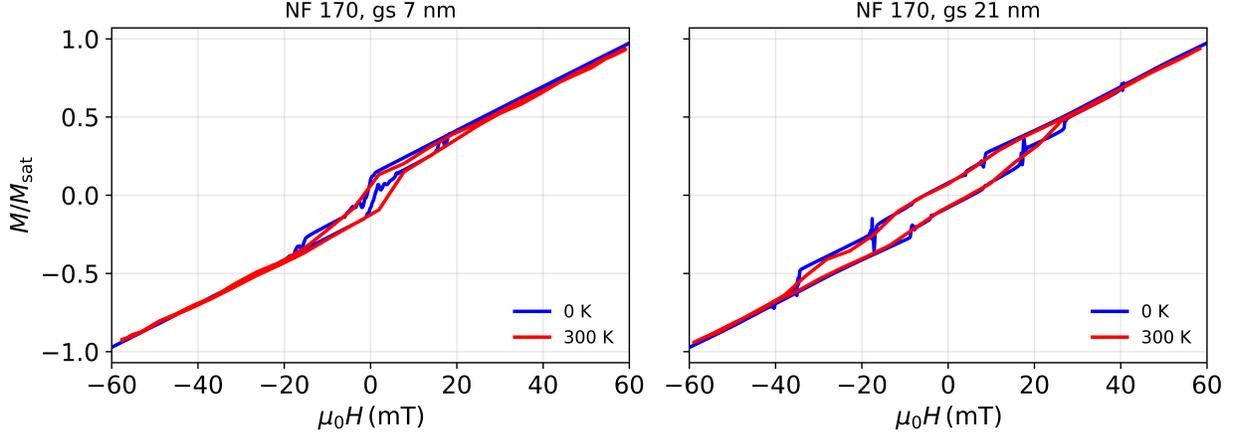}}
    \caption{Hysteresis loops calculated for $d = 170$ nm NFs with $g_{s} = 7$ and 21 nm, respectively, at $T = 0~\mathrm{K}~\mathrm{and}~300~\mathrm{K}$. Driving field parameters are $f = 300~\mathrm{kHz}$ and $h_{AC}=\pm 49.33~\mathrm{kA/m}$. No significant changes in the coercivity are observed.}
    \label{AC_loops_temperature}
\end{figure*}

To assess the impact of finite temperature on the extracted energy losses, we directly compared hysteresis loops obtained from simulations performed at $T = 0~\mathrm{K}$ and $T = 300~\mathrm{K}$, with AC field excitation of $f = 300~\mathrm{kHz}$ and maximum field amplitude of $h_{AC} = \pm 49.33~\mathrm{kA/m}$. Figure~\ref{AC_loops_temperature} presents averaged loops (10 realizations) for NFs of $d = 170~\mathrm{nm}$, combined with grain sizes of $g_{s} = 7~\mathrm{nm}$ and $21~\mathrm{nm}$. In all cases, the loops display a similar qualitative behaviour: in the loops at $T = 0~\mathrm{K}$, the magnetization exhibits some jumps, revealing the pinning of the vortex core at the inter-grain boundaries while the magnetization reversal takes place. Those jumps are suppressed at $T = 300~\mathrm{K}$, where the thermal fluctuations provide an extra energy to overcome the pinning energy barriers, resulting in a smoother magnetization reversal with no pinning. However, the loop coercivity remains almost constant at both temperature values, constraining the effect of thermal fluctuations to the magnetization dynamics, yet with little to negligible effect on the macroscopic magnetic energy stored per cycle. This demonstrates that the dominant dissipation mechanism is field-driven, with minimal influence of thermal effects at the proven fields. \newline

\section{Simulations for larger NFs: $d =$ 220 nm and 250 nm}
\begin{figure*}
    \centering
    \resizebox{1\columnwidth}{!}{\includegraphics{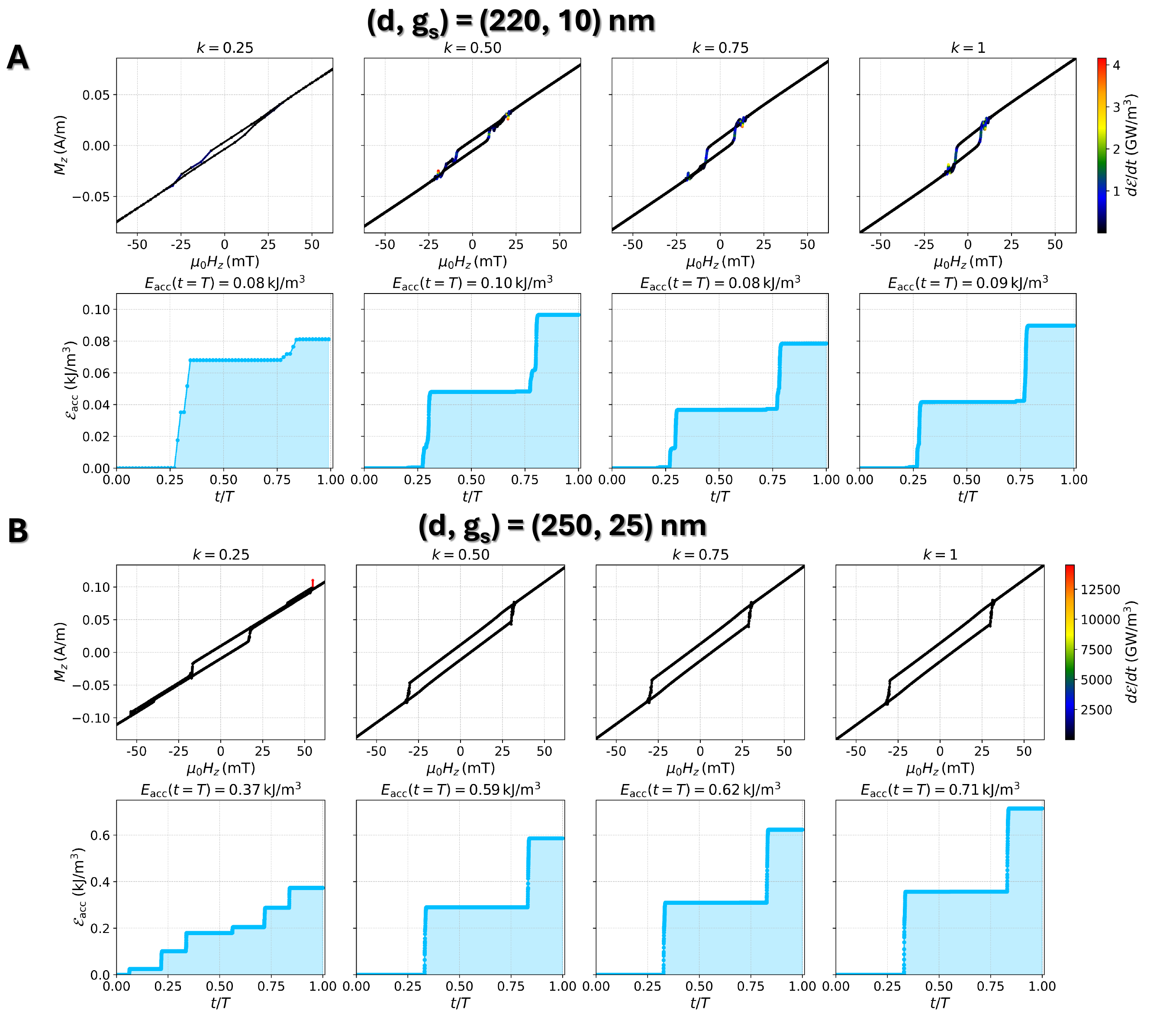}}
    \caption{Magnetization as a function of applied field and corresponding released energy $\mathcal{E}(t)$ over one AC-field period $T = 1/300~\mathrm{ms}$ for NFs with diameters  $d=220~\mathrm{nm}$ and $g_{s}=10~\mathrm{nm}$ (\textbf{A}), and $d=250~\mathrm{nm}$ and $g_{s}=25~\mathrm{nm}$ (\textbf{B}). The magnetization traces are colour-coded by the instantaneous heat generation, with red highlighting the points of maximum dissipation, coinciding with the jumps observed in the magnetization consequence of the reversal, \textit{i.e.} moments of maximized torque. The step-like increases in $\mathcal{E}(t)$ coincide with these dissipation bursts. Inter-grain exchange is fixed to $k\times A$, with $k = 1.00$.}
    \label{250_25_220_10_primer}
\end{figure*}

\begin{figure*}
    \centering
    \resizebox{1\columnwidth}{!}{\includegraphics{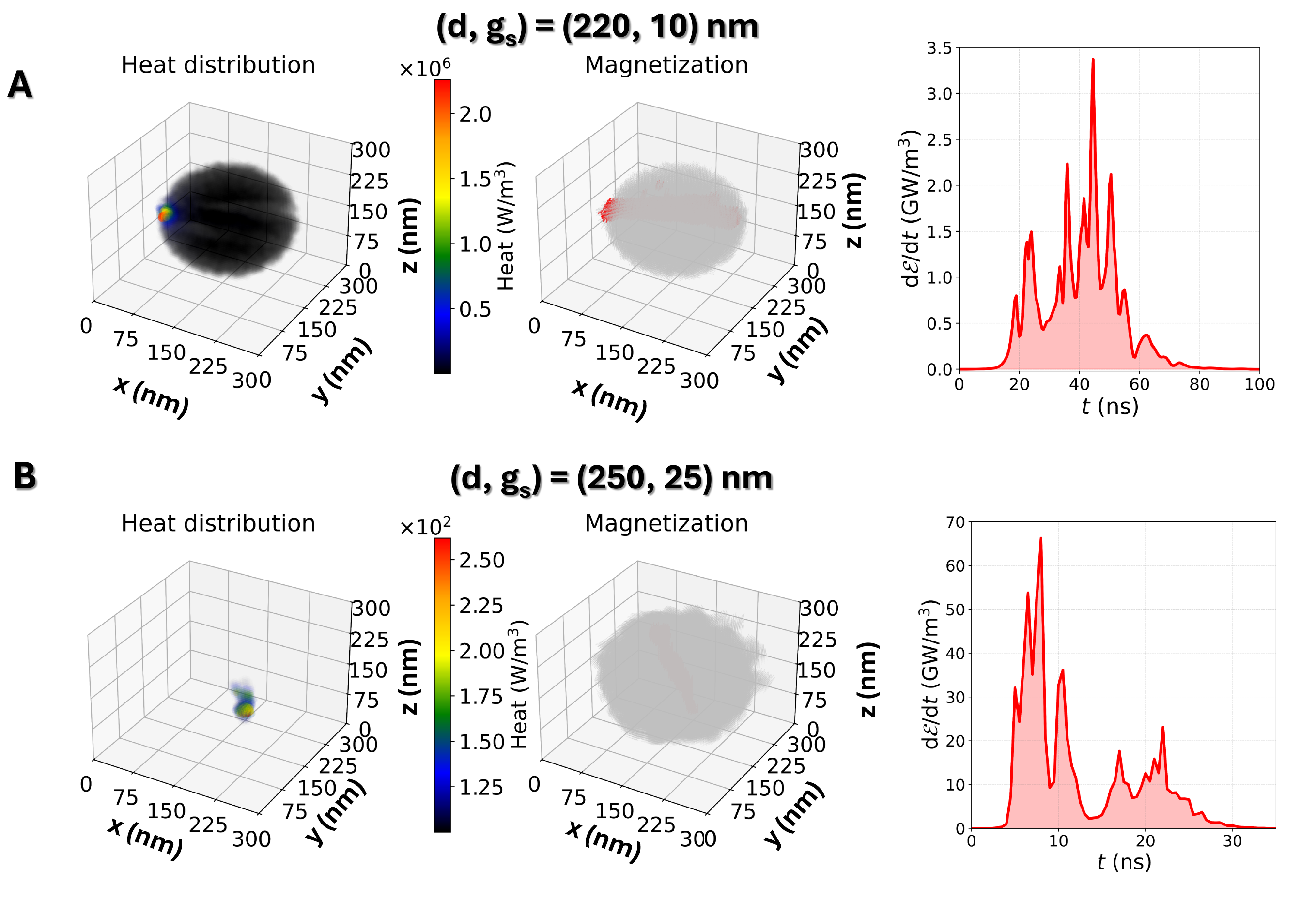}}
    \caption{3D heat and magnetization mappings, along with the instantaneous heat release $d\mathcal{E}/dt$ vs. $t$, for \textbf{(A)} $(d, g_{s}) = (220,10)$ nm, and \textbf{(B)} $(d, g_{s})=$(250,25) nm, respectively. Inter-grain exchange is fixed to $k\times A$, with $k = 1.00$.}
    \label{250_25_220_10_second}
\end{figure*}

Figures~\ref{250_25_220_10_primer} and \ref{250_25_220_10_second} include the results corresponding to $(d, gs) = (220, 10)~\mathrm{and}~(250, 25)$ NFs. Akin to the simulations shown in the main text, the heat release is concentrated at the extremes of the magnetization reversal. The energy accumulated per cycle a factor 6 larger for the larger grains compared to the smaller counterparts, in agreement with the simulations for $d = 170$ nm and the experimental results. The heat release of the NFs with smaller grains (220, 10) is concentrated spatially around the vortex core, extending over a longer period of time ($t\sim 100$ ns) compared to the realizations with larger grains (250, 25), which deliver the heat at a shorter time windows ($t\sim 25$ ns).  \newline

\section{Heat generation of a single domain spherical nanoparticle}
\begin{figure*}
    \centering
    \resizebox{1\columnwidth}{!}{\includegraphics{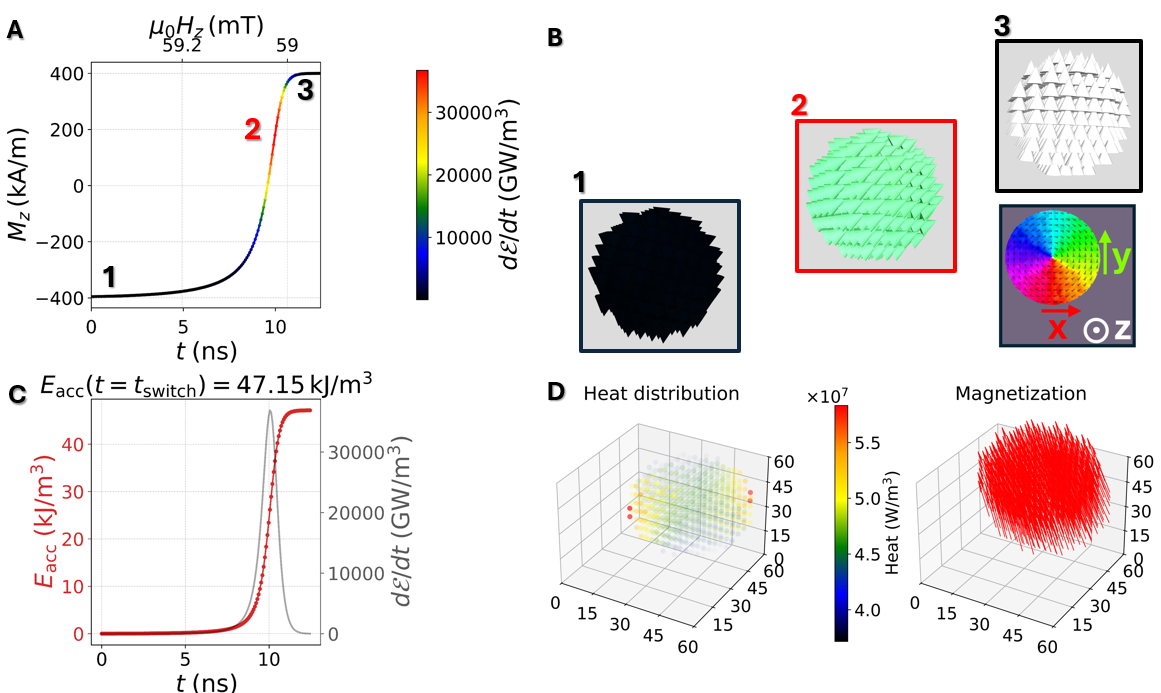}}
    \caption{Single-domain 50 nm diameter nanosphere. \textbf{(A)} Magnetization as a function of applied field (top axis) and time (bottom axis). The magnetization traces are colour-coded by the instantaneous heat generation, with red highlighting the points of maximum dissipation. \textbf{(B)} Snapshots of three representative time steps of the magnetization reversal: 1 and 3 correspond to the magnetization along $z$-axis, 2 to the magnetization along the $y$ direction. We have included the colour-coded wheel on the right side. \textbf{(C)} Instantaneous heat release (grey) and energy accumulated (red) for the reversal process, which happens within 5 nanoseconds. \textbf{(D)} temperature and magnetization mappings corresponding to the moment of maximum heat release. The heat is spread along the whole nanoparticle.}
    \label{single_domain}
\end{figure*}

Figure~\ref{single_domain} \textbf{(A)} includes the magnetization reversal as a function of the applied magnetic field and time. Regions marked as 1 and 3 correspond to the magnetization purely oriented along the $z$--axis (see \textbf{(B)}), whereas region 2 marks the area where the magnetization points perpendicular to the field direction. This region corresponds to the heat release, included in \textbf{(C)} as instantaneous heat release (gray line) and energy accumulated within the probed period. In \textbf{(D)}, the heating and magnetization mappings are shown for the moment of maximum heat release. Contrary to the vortex-like textures discussed in the work, the heat release of single-domain particles is homogeneous. Note the difference to Fig. 4 of the main text, where the heat release is purely concentrated along the vortex core direction, and extended over tens of nanoseconds, whereas in the single-domain case, the heat is released an order of magnitude faster ($t\sim 5~\mathrm{ns}$).\newline

\clearpage
\bibliography{main}
